\documentclass[apjl,twocolumn]{openjournal}
\usepackage{natbib} 
\usepackage[dvipsnames]{xcolor}
\usepackage{aas_macros} 
\usepackage{amssymb}
\usepackage{amsmath}
\usepackage{threeparttable}
\usepackage[title]{appendix}
\usepackage{hyperref}	
\hypersetup{colorlinks=true,linkcolor=blue,citecolor=blue,filecolor=blue,urlcolor=blue}
\usepackage[caption=false]{subfig}
\usepackage{soul}                          

\begin{document}
\title[The Nebular Continuum at High-Redshift]{21 Balmer Jump Street: The Nebular Continuum at High Redshift and Implications for the Bright Galaxy Problem, UV Continuum Slopes, and Early Stellar Populations\vspace{-15mm}}
\author{Harley Katz$^{1}$\thanks{$^*$E-mail: \href{mailto:harleykatz@uchicago.edu}{harley.katz@physics.ox.ac.uk}}, 
Alex J. Cameron$^{2}$, 
Aayush Saxena$^{2}$,
Laia Barrufet$^{3}$,
Nicholas Choustikov$^{2}$,
Nikko J. Cleri$^{4,5,6}$,
Anna de Graaff$^7$,
Richard S. Ellis$^8$,
Robert A. E. Fosbury$^9$,
Kasper E. Heintz$^{10,11,12}$,
Michael Maseda$^{13}$,
Jorryt Matthee$^{14}$, 
Ian McConachie$^{13,15}$,
Pascal A. Oesch$^{12,10,11}$
}

\affiliation{$^{1}$Department of Astronomy \& Astrophysics, University of Chicago, 5640 S Ellis Avenue, Chicago, IL 60637, USA}
\affiliation{$^{2}$Department of Physics, University of Oxford, Denys Wilkinson Building, Keble Road, Oxford, OX1 3RH, UK}
\affiliation{$^{3}$Institute for Astronomy, University of Edinburgh, Royal Observatory, Edinburgh, EH9 3HJ}
\affiliation{$^{4}$Department of Astronomy and Astrophysics, The Pennsylvania State University, University Park, PA 16802, USA}
\affiliation{$^{5}$Institute for Computational and Data Sciences, The Pennsylvania State University, University Park, PA 16802, USA}
\affiliation{$^{6}$Institute for Gravitation and the Cosmos, The Pennsylvania State University, University Park, PA 16802, USA}
\affiliation{$^{7}$Max-Planck-Institut für Astronomie, Königstuhl 17, D-69117, Heidelberg, Germany}
\affiliation{$^{8}$Department of Physics and Astronomy, University College London, Gower Street, London WC1E 6BT, UK}
\affiliation{$^{9}$European Southern Observatory, Karl-Schwarzschild-Strasse 2, 85748 Garching, Germany}
\affiliation{$^{10}$Cosmic Dawn Center (DAWN), Denmark}
\affiliation{$^{11}$Niels Bohr Institute, University of Copenhagen, Jagtvej 128, DK-2200 Copenhagen
N, Denmark.}
\affiliation{$^{12}$Department of Astronomy, University of Geneva, Chemin Pegasi 51, 1290 Versoix, Switzerland.}
\affiliation{$^{13}$Department of Astronomy, University of Wisconsin–Madison, 475 N. Charter St., Madison, WI 53706 USA}
\affiliation{$^{14}$Institute of Science and Technology Austria (ISTA), Am Campus 1, 3400 Klosterneuburg, Austria}
\affiliation{$^{15}$Department of Physics and Astronomy, University of California, Riverside, 900 University Avenue, Riverside, CA 92521, USA}

\begin{abstract}
We study, from both a theoretical and observational perspective, the physical origin and spectroscopic impact of extreme nebular emission in high-redshift galaxies. The nebular continuum, which can appear during an extreme starburst, is of particular importance as it tends to redden UV slopes and has a significant contribution to the UV luminosities of galaxies. Furthermore, its shape can be used to infer the gas density and temperature of the interstellar medium. First, we provide a theoretical background, showing how different stellar populations (SPS models, initial mass functions (IMFs), and stellar temperatures) and nebular conditions impact observed galaxy spectra. We demonstrate that, for systems with strong nebular continuum emission, 1) UV fluxes can increase by up to 0.7~magnitudes (or more in the case of hot/massive stars) above the stellar continuum, which may help reconcile the surprising abundance of bright high-redshift galaxies and the elevated UV luminosity density at $z\gtrsim10$, 2) at high gas densities, UV slopes can redden from $\beta\lesssim-2.5$ to $\beta\sim-1$, 3) observational measurements of $\xi_{\rm ion}$ are gross underestimates, and 4) UV downturns from two-photon emission can masquerade as damped Ly$\alpha$ systems. Second, we present a dataset of 58 galaxies observed with NIRSpec on JWST at $2.5<z<9.0$ that are selected to have strong nebular continuum emission via the detection of the Balmer jump. Five of the 58 spectra are consistent with being dominated by nebular emission, exhibiting both a Balmer jump and a UV downturn consistent with two-photon emission. For some galaxies, this may imply the presence of hot massive stars and a top-heavy IMF. We conclude by exploring the properties of spectroscopically confirmed $z>10$ galaxies, finding that UV slopes and UV downturns are in some cases redder or steeper than expected from SPS models, which may hint at more exotic (e.g. hotter/more massive stars or AGN) ionizing sources.
\end{abstract}

\section{Introduction}
The successful launch of JWST \citep{Garnder2006} has opened a new frontier to study galaxy formation at the earliest cosmic epochs. Understanding the physical properties of these early galaxies, how they differ from those in the low-redshift Universe, and how they impacted subsequent galaxy evolution remain key open goals of modern extragalactic astronomy. One of the primary advantages of JWST compared to HST is its spectroscopic capabilities in the rest-frame UV and optical at high-redshift. This tool can be used not only to confirm the redshifts estimated from photometry, but also to obtain deep insights into the stellar populations, star formation histories, and the conditions of the interstellar medium (ISM) of galaxies in the early Universe \citep[e.g.][]{Sanders2023b,Cameron2023_neb}.

Numerous galaxies have now been spectroscopically confirmed at $z>10$ \citep[e.g.][]{CurtisLake2023,Wang2023,Haro2023,Carniani2024} and have been shown to exhibit a diverse set of properties. For example GNz11 \citep{Bunker2023} and GHZ2 \citep{Castellano2024,Zavala2024} are characterized by extreme UV emission lines, including those rarely seen at lower redshifts (e.g. N~{\small IV}]~$\lambda$1486), while others exhibit much lower EW (and undetected) UV lines \citep[e.g.][]{Carniani2024,CurtisLake2023}. Compilations of high-redshift galaxies \citep[e.g.][]{Heintz2024,Roberts-Borsani2024} show that UV slopes can range from close to $-3$, which would be consistent with the intrinsic values of massive stars, to redder than $-2$, which is typically reconciled by assuming dust attenuation, clearly indicating a difference in ISM properties or stellar populations. With JWST, we are seeing unprecedented spectral features at high-redshift.

More generally, the photometric detection and spectroscopic confirmation of numerous bright galaxies at $z\gtrsim10$ as well as the elevated UV luminosity density at similar redshifts is in tension with most theoretical models of high-redshift galaxy formation \citep[e.g.][although c.f. \citealt{Willott2023}]{Finkelstein2023_b,Harikane2024,Leung2023,Chemerynska2023}. Various physical mechanisms have been proposed to explain this overabundance of bright (${\rm M_{UV}}\gtrsim-19$) galaxies such as:
\begin{enumerate}
\item an increased scatter in the relation between dark matter halo mass and UV luminosity compared to the lower-redshift Universe, possibly driven by bursty star formation \citep[e.g.][]{Ren2019,Mason2023,Shen2023,Sun2023,Kravtsov2024},
\item an increased star formation efficiency at high redshift \citep[e.g.][]{McCaffrey2023}, potentially caused by weak feedback \citep[e.g.][]{Dekel2023},
\item a lack of dust in massive galaxies due to strong radiation pressure \citep{Ferrara2023},
\item or a change in the mass-to-light ratio of early stellar populations, possibly due to a top-heavy IMF (e.g. \citealt{Yung2024}, although c.f. \citealt{Cueto2024}).
\end{enumerate}
These solutions address the problem by either changing the observed mass-to-light ratio or forming stars more efficiently. The true origin of the bright galaxy problem and excess UV luminosity density is likely a combination of many physical effects. For example, nearly all high-resolution numerical simulations of high-redshift galaxies that model cold gas in the ISM predict that star-formation is bursty \citep[e.g.][]{Ma2018,spdrv1,Pallottini2022}. However, as this scatter is mass-dependent \citep[e.g.][]{Gelli2024,Kravtsov2024} it remains an open question whether the level of SFR burstiness generically expected in galaxy formation models is sufficient to produce the level of UV luminosity variability required to match observations \citep[e.g.][]{Pallottini2023}. 

Numerous 1D \citep[e.g.][]{Omukai2005} and 3D models \citep[e.g.][]{Chon2022} predict that the stellar initial mass function (IMF) becomes more top-heavy at lower metallicities and in the early Universe when the cosmic microwave background (CMB) temperature is higher. A more top-heavy IMF is a generic prediction in the extreme limit of metal-free star formation at high redshift \citep[e.g.][]{Bromm2002,Abel2002,Hirano2014}. The mass-to-light ratio is lower at young stellar ages for IMFs with flatter high-mass slopes, which would lead to an increase in UV luminosity at fixed star formation rate. 

\begin{figure}
\centering
\includegraphics[width=0.45\textwidth]{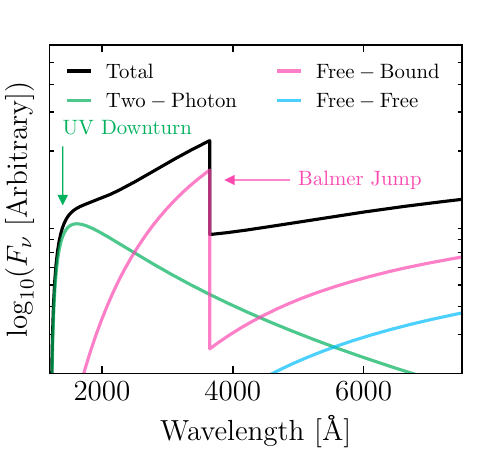}
\caption{Spectral shape of the three components of the nebular continuum as a function of wavelength. We show the shape of the nebular continuum for gas at $2\times10^4$~K and an electron density of $10^2\ {\rm cm^{-3}}$ assuming Case~B recombination. Annotated are the locations of the Balmer jump caused by free-bound emission and the UV downturn from two-photon emission.}
\label{fig:cartoon}
\end{figure}

What has perhaps been considered less in the literature is the role of the nebular continuum, which is observed to become important in lower mass, low-metallicity environments \citep[e.g.][]{Izotov2011}. When ionizing photons are absorbed by gas they can be reprocessed into continuum emission through three different processes: free-free, free-bound, and two-photon\footnote{Note that throughout this work we will only consider the nebular continuum contribution as we will analyze spectroscopic data. For photometric data, line emission can also contribute to the inferred luminosity integrated over a photometric passband.}. For a typical star-forming galaxy, nebular continuum emission is generally subdominant at rest-frame wavelengths of 1500~\AA, where UV luminosities are typically measured. However, strong nebular continuum emission has now been detected in both low \citep[e.g.][]{Guseva2006,Guseva2007} and high-redshift galaxies \citep[e.g.][]{Cameron2023_BJ,Roberts-Borsani2024,Welch2024} undergoing bursts of star formation. 

Free-bound emission is arguably the most easily recognizable due to the Balmer jump, which is a sharp discontinuity at rest-frame 3645~\AA\ (see the pink line in Figure~\ref{fig:cartoon}). Two-photon emission, particularly the $2s\rightarrow1s$ transition of neutral hydrogen peaks very close to 1500~\AA~(in $f_{\lambda}$) and smoothly drops towards both shorter and longer wavelengths, going to zero at 1216~\AA\ (see the green line in Figure~\ref{fig:cartoon}). In addition to being emitted by H~{\small II} regions, two-photon emission can also arise from warm cooling flows that may be present in the high-redshift Universe \citep[e.g.][]{Dijkstra2009}. Two-photon emission is rarely directly detected because for Population II star formation, the power-law stellar continuum increases towards the blue end of the spectrum and typically outshines the nebula at short wavelengths.  However, in low-metallicity environments where H~{\small II} region temperatures are elevated, Case~B departures can occur which increases both the two-photon and Ly$\alpha$ emission via collisional excitation \citep[e.g.][]{Raiter2010,Ribas2016}. 

In contrast to two-photon emission, free-bound emission increases towards longer wavelengths and can thus outshine the stellar emission. Hence it can be directly detected in the rest-frame optical. The key aspect of the nebular continuum is that it should only become important when galaxies are undergoing a burst of star formation. The natural UV magnitude fluctuations driven by deviations from the star-forming main-sequence can be further enhanced by continuum emission from the gas. Identifying a sample of high-redshift galaxies with strong nebular continuum emission may provide an ideal test-bed for elucidating the physical mechanisms that drive extreme bursts of star formation, which may explain the over-abundance of bright high-redshift galaxies and the excess UV luminosity density. 

Unfortunately, a sample of high-redshift galaxies with detected nebular continuum has yet to be compiled. Only one individual galaxy with a Balmer jump has been reported at $z>5$ \citep{Cameron2023_BJ} and this same feature has appeared in a stack of $z\sim6$ Ly$\alpha$ emitters \citep{Roberts-Borsani2024}.

The lack of reported Balmer jumps is perhaps surprising because high-resolution simulations of early galaxies generically predict bursty star formation histories such that the SFRs can fluctuate by multiple orders of magnitude on tens of Myr timescales \citep[e.g.][]{Ma2018,spdrv1,Pallottini2022}, which is exactly the scenario where the nebular continuum is expected to be strong. SED fitting of the photometry of high-redshift galaxy candidates often predicts extreme nebular emission \citep[e.g.][]{Endsley2023,Bradac2024}. Moreover, the flux-limited nature of high-redshift surveys preferentially biases observations towards galaxies that are highly star-forming, which is when the nebular continuum should begin to appear. The spectral resolution of the NIRSpec prism and sensitivity of the grating may also impact our ability to detect the Balmer jump if present. As the prevalence and characteristics of the nebular continuum at high-redshift are poorly known, its importance for the luminosities of early galaxies remains to be quantified. 

In this work, we aim to better understand the physics of galaxies undergoing extreme bursts of star formation by focusing on the spectral impact of extreme nebular continuum emission and the underlying physics that may be driving its observability. We begin in Section~\ref{sec:background} by providing a theoretical background for how nebular emission modifies the observational properties of star-bursting galaxies, focusing on changes to the properties of the underlying stellar populations and ISM conditions. In Section~\ref{sec:observations}, we test the theoretical results in the context of a newly compiled sample of 58 spectroscopically confirmed high-redshift galaxies that show Balmer jumps. Finally, in Sections~\ref{sec:discuss}~and~\ref{sec:conclusion}, we discuss the implications our findings in the context of $z>10$ galaxies and present our conclusions.

\section{Theoretical Background}
\label{sec:background}
We begin by providing a theoretical overview of the expected nebular continuum contribution of different stellar populations to the observed spectra of star-bursting galaxies and how this emission impacts our ability to infer the underlying properties of these same galaxies.

\begin{figure*}
\centering
\includegraphics[width=\textwidth]{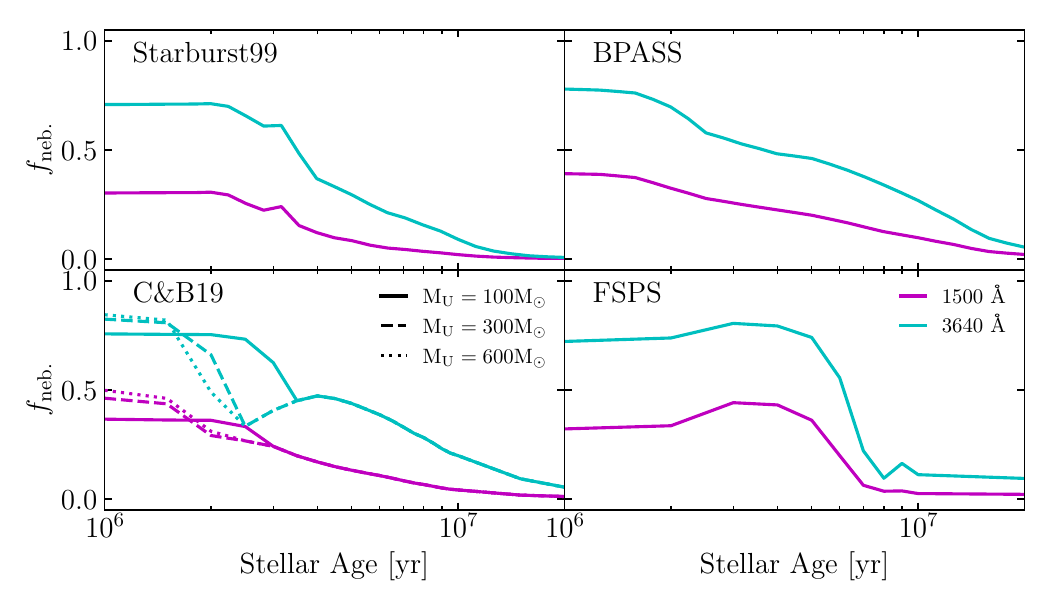}
\caption{Nebular continuum fraction of the total SED at either 1500~\AA~(magenta), where the two-photon continuum becomes important, or at 3640~\AA~(cyan), where the free-bound emission reaches its maximum. We show results for four different SPS models as a function of stellar age. For the C\&B19 models, different line styles represent different upper mass limits for the IMF. Note that all models shown assume an H~II region density of $10^3\ {\rm cm^{-3}}$.}
\label{fig:sps_fneb}
\end{figure*}

\subsection{Spectral synthesis modeling}
SEDs of galaxies including the nebular continuum are computed with {\small CLOUDY} version C23 \citep{Chatzikos2023}. Unless otherwise specified, models assume a slab configuration with a dimensionless ionization parameter of $\log_{10}U=-2.0$, a metallicity of $1\%$ solar by scaling down abundance ratios from \cite{Asplund2009}, and a hydrogen gas density of $10^3~{\rm cm^{-3}}$, consistent with that of galaxies in the early Universe \citep[e.g.][]{Isobe2023}. Calculations are stopped when the electron fraction reaches 1\%\footnote{We consider this to be the edge of an ionization bounded nebula. Adopting a density bounded model would reduce the contribution of nebular emission with respect to the stellar continuum as the escape fraction will be non-zero. Moreover, extending the calculation beyond the Strömgren sphere will not impact our results unless the gas is hot enough so that it can be collisionally excited. This possibility is further discussed below.}. The exact details of the models have little impact on our conclusions. We consider five different stellar population synthesis (SPS) models, {\small Starburst99} \citep{Leitherer1999}\footnote{\href{https://www.stsci.edu/science/starburst99/docs/default.htm}{https://www.stsci.edu/science/starburst99/docs/default.htm}}, {\small BPASS}~v2.2.1 \citep{Stanway2018}\footnote{\href{https://bpass.auckland.ac.nz/9.html}{https://bpass.auckland.ac.nz/9.html}}, {\small FSPS} \citep{Conroy2009,Conroy2010}\footnote{\href{https://github.com/cconroy20/fsps}{https://github.com/cconroy20/fsps}}, {\small C\&B 2019} models\footnote{\href{http://www.bruzual.org}{http://www.bruzual.org}}, and individual, metal-poor, massive star models of \cite{Larkin2023}\footnote{\href{https://atmos.ucsd.edu}{https://atmos.ucsd.edu}}. Within each subset of models, we choose the SEDs that have the closest metallicity to the fiducial value of 1\% that we adopt for the gas\footnote{Note that we do not assume any $\alpha$ enhancement in stellar SEDs which is expected both at low metallicities and at high redshifts \citep[e.g.][]{Kobayashi2020,Cullen2021}.}. All models assume an instantaneous burst of star formation and we consider stellar ages up to 20~Myr. Beyond this, the ionizing output is so low that the nebular emission is negligible. Note that the upper mass limits for all of our chosen SPS models differ. The fiducial upper mass cutoffs are 120~${\rm M_{\odot}}$, 300~${\rm M_{\odot}}$, and 300~${\rm M_{\odot}}$ for {\small Starburst99}, {\small BPASS}\footnote{For the BPASS models, we adopt a Kroupa IMF \citep{Kroupa2001}.}, and {\small FSPS}\footnote{For the FSPS models we adopt a Chabrier IMF \citep{Chabrier2003} adopting the MIST isochrones \citep{Choi2016} and MILES spectra \citep{Vazdekis2010}.}, respectively. For the {\small C\&B19} models\footnote{We have used a Chabrier IMF \citep{Chabrier2003} for the C\&B19 models.}, we consider three different upper mass limits of 100~${\rm M_{\odot}}$, 300~${\rm M_{\odot}}$, and 600~${\rm M_{\odot}}$.

The output from these models are the transmitted stellar emission as well as the continuum and line emission from the gas. It should be noted that in real galaxies, the turbulent nature of H~{\small II} regions can modify the gas emission from expectations of spherical models \citep[e.g.][]{Jin2022} and interpretation biases can arise if multiple H~{\small II} regions with different properties combine to produce the spectrum of a galaxy \citep[e.g.][]{Cameron2023}. Nevertheless, the 1D models used here provide both quantitative and qualitative insights into how galaxy spectra may change under different conditions under a restrictive set of assumptions. 

\subsection{How strong can the nebular continuum be?}
We first address the question of how much the nebular continuum can contribute to the UV luminosity of a galaxy.

\begin{figure}
\centering
\includegraphics[width=0.45\textwidth]{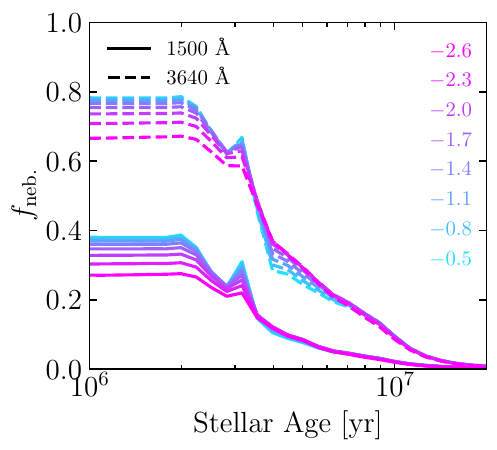}
\includegraphics[width=0.45\textwidth]{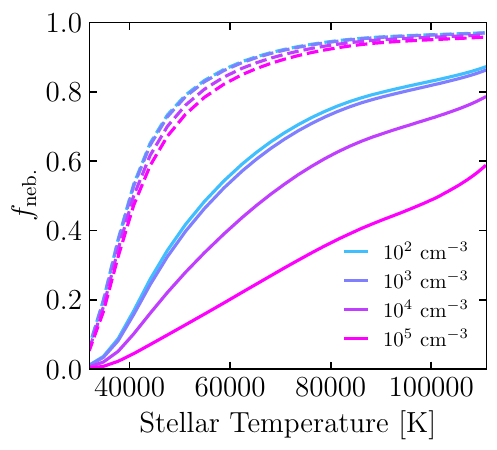}
\caption{Contribution of the nebular continuum to the total SED at either 1500~\AA~(solid), where the two-photon continuum becomes important, or at 3640~\AA~(dashed), where the free-bound emission reaches its maximum. We show results for Starburst99 models assuming a gas density of $10^3~{\rm cm^{-3}}$ with different stellar initial mass functions where the upper-mass slope is varied as listed in the legend (top) as well as for low-metallicity massive star models of different stellar temperatures (bottom) where we vary the gas density of the nebula. For models with varying stellar temperature, we only consider the zero-age main-sequence spectra.}
\label{fig:imf_temp_fneb}
\end{figure}

\subsubsection{Commonly Used SPS models}
\label{sec:standard_sps}
In Figure~\ref{fig:sps_fneb} we show the fraction of the total 1500~\AA\ or 3640~\AA~continuum emission from the galaxy comes from the nebula as a function of stellar age for four SPS models. For young stellar ages, $\sim1$~Myr, all models predict that the nebula contributes at least 30\% to the 1500~\AA\ luminosity, which corresponds to brightening the galaxy by $\sim0.4$ magnitudes. The nebular continuum contribution can be $>70\%$ in the rest-frame optical. All SPS models predict that after $\sim10$~Myr, the nebular contribution at 1500~\AA\ is negligible, a consequence of the massive stars evolving off the main sequence and the lower ionizing photon production efficiencies of less massive stars. The timescale over which the nebular continuum remains important is slightly longer for the free-bound emission due to the fact that it is stronger with respect to the stellar continuum compared to two-photon emission. We note that burst models maximize the nebular contribution to each SSP and the nebular continuum contribution for realistic SFHs will be explored in Section~\ref{sec:cavs_and_such}.

While the general trends between the SPS curves are similar, in detail the shapes of the curves and the maximum nebular contributions are different. For example the 1500~\AA\ nebular contribution of the {\small BPASS} models peak at $\sim40\%$ at 1~Myr, much higher than the {\small Starburst99} models, due to the increased ionizing photon production efficiencies caused by binary stellar evolution. The {\small C\&B19} models with maximum stellar masses of 300~M$_{\odot}$ and 600~M$_{\odot}$ can reach a 1500~\AA\ nebular contribution as high as 50\% at similarly young stellar ages, which corresponds to a brightening the UV luminosity by 0.75~magnitudes. In contrast to the other three SPS models, the nebular contribution of the {\small FSPS} models peaks between $3-5$~Myr at nearly 45\%, which results from different assumptions regarding Wolf-Rayet stars. 

\subsubsection{A top-heavy IMF}
Numerous theoretical models predict that the stellar IMF becomes more top-heavy at high redshift due to lower gas metallicities and an elevated CMB temperature \citep[e.g.][]{Chon2022,Bate2023}. This has two important effects. First, as the upper-mass IMF slope becomes flatter\footnote{Here we only consider modifications to the IMF where the power-law slope of the high-mass end is varied. However, the shape of the IMF may be fundamentally different and deviate from a power-law at high-redshift \citep[e.g.][]{Chon2022}. For our work, the important parameter is the fraction of mass in high-mass stars with respect to that at lower masses rather than the exact shape.}, at young stellar ages, more UV luminosity is produced per unit star formation, which would help explain the bright galaxy problem by decreasing the mass-to-light ratio at young ages. Second, the ionizing photon production per unit star formation also increases as the IMF becomes more biased towards high-mass stars. To test the impact of a varying IMF, we adopt {\small Starburst99} models\footnote{{\small Starburst99} models were adopted because it is trivial to vary the properties of the IMF, in contrast to some of the other SPS models. Our primary conclusions likely generalize to the other SPS models.}, where we have systematically varied the upper-mass slope, $\alpha$, from $-2.6$ (i.e. more bottom heavy than the canonical $-2.3$) to $-0.5$ (where nearly all of the mass is locked in massive stars). We define the IMF such that:
\begin{equation}
\xi(m) \propto
\begin{cases} 
    m^{-1.3} & m \leq 0.5\ {\rm M_{\odot}} \\
   m^{\alpha} &  m > 0.5\ {\rm M_{\odot}}
\end{cases}
\end{equation}
The lower and upper mass limits of the IMF are set to 0.08~${\rm M_{\odot}}$ and 120~${\rm M_{\odot}}$, respectively.

In the top panel of Figure~\ref{fig:imf_temp_fneb}, we show the fractional nebular contribution at 1500~\AA\ and 3640~\AA\ as a function of stellar age. The general trend of the nebular contribution is the same as above -- the peak in nebular emission comes at young stellar ages and begins to subside after a few Myr, with a small secondary peak from Wolf-Rayet stars. Unsurprisingly, a more top-heavy IMF results in an increased fractional nebular contribution to the 1500~\AA\ luminosity. While the fiducial model predicts that $\sim30\%$ of the 1500~\AA\ luminosity at an age of 1~Myr comes from nebular emission, by flattening the upper-mass IMF slope to $-0.5$, the nebular contribution increases to $\sim38\%$ at the same stellar age. This increase is rather modest, especially compared to the total increase in UV luminosity which is more than a factor of ten for an upper-mass IMF slope to $-0.5$ (see below).

It is straightforward to understand why extreme IMF variations have only a minimal impact on the nebular continuum fraction at 1500~\AA. A massive O star has temperature of $\sim40,000$~K \citep{Pecaut2013}, which according to Wien's law has a spectral peak at $\sim725$~\AA. For a standard IMF, these stars dominate the ionizing output as going slightly lower in temperature pushes the peak redward of the hydrogen ionizing wavelength. Since hotter blackbodies are brighter at all wavelengths, the most massive stars also contribute significantly to the 1500~\AA\ UV luminosity, despite both their lower numbers and their spectral peak at shorter wavelengths. The absolute maximum nebular contribution is then that which comes from the most massive star in the IMF. Since it is these stars that provide both the ionizing photons and a significant amount of the 1500~\AA\ luminosity, flattening the upper-mass slope of the IMF (i.e. making it more ``top-heavy'') does not substantially increase the fractional nebular contribution because the upper limit to this value is set by the most massive star in the IMF.

\subsubsection{Hotter and more massive stars}
\label{sec:hot_massive}
We have shown that for an IMF with a fixed upper mass, the fractional nebular contribution at 1500~\AA\ does not increase substantially if the upper-mass IMF slope flattens from $-2.3$ to $-0.5$. However, increasing the maximum mass sampled by the IMF represents a possible alternative. This was already hinted at in Section~\ref{sec:standard_sps} where we showed for the {\small C\&B19} models that increasing the maximum mass from 100~M$_{\odot}$ to 600~M$_{\odot}$ increased the fractional nebular contribution at both 1500~\AA\ and 3640~\AA. 

To explore this effect further, we adopt the low-metallicity, massive star models of \cite{Larkin2023} which are available for a range of stellar temperatures. We assume that these are generally representative of low-metallicity massive stars in the early Universe for which there is limited public data of stellar atmospheres. The effect on nebular contribution strength due to very massive stars was recently discussed in \cite{Schaerer2024b}. Our work differs from \cite{Schaerer2024b} in that they chose a very specific model for massive stars at a metallicity that is higher than typically measured for high-redshift galaxies and hence the effective temperatures of those models are lower than considered here. Moreover, they apply a more simplistic model for the nebular continuum by assuming a constant temperature, which both ignores the temperature structure of the nebula and neglects potential Case~B departures that may become important at low metallicity \citep[e.g.][]{Raiter2010,Ribas2016}. 

In the bottom panel of Figure~\ref{fig:imf_temp_fneb} we show the fractional nebular contribution at 1500~\AA\ and 3640~\AA\ as a function of stellar temperature. While at a temperature of 40,000~K, the nebular fraction at 1500~\AA\ is only $\sim14\%$, for stars with $T=100,000$~K, nebular emission represents 80\% of the total 1500~\AA\ UV luminosity. For these hot stars, the UV luminosity could then be brightened by 1.75~magnitudes, purely due to nebular emission.

This prediction is not unique to our work. In fact, a generic characteristic of hot star models is that they have nebular dominated spectra across nearly the entire UV and optical \citep[e.g.][]{Panagia2002,Schaerer2003,Trussler2023}, and \cite{Raiter2010} have already shown that the nebular continuum can dominate the 1500~\AA\ UV luminosity in the context of Pop.~III stars under various assumptions of IMF. \cite{Raiter2010} also pointed out that at low metallicities, one expects departures from standard Case~B assumptions due to the fact that the $2s$ and $2p$ states can be collisionally populated. Case~B departures have the effect of enhancing the two-photon emission (and Ly$\alpha$), both due to the fact that the collisionally excited atom can fluoresce down to the ground state and that in the excited state, lower energy photons are needed to ionize the gas, although this latter effect is likely subdominant \citep{Ribas2016}. Case~B departures due to collisional excitation will also increase the intrinsic ratio of H$\alpha$/H$\beta$ to values $>2.86$. 

\subsubsection{Other physics that impacts the nebular contribution}
\paragraph{Cooling Flows}
Just as collisions impact the level populations within H~{\small II} regions, gas accreting onto haloes can undergo the same process as long as the virial temperature of the halo is high enough ($\gtrsim10^4$~K) \citep[e.g.][]{Fardal2001,FG2010}. This process then cools the gas and is particularly efficient in the temperature range $15,000-20,000$~K. Numerical simulations predict that cold-mode accretion, where cooling is strongest is much more prevalent at high-redshift compared to the local Universe \citep{Keres2005}. A nebular-only spectrum dominated by Ly$\alpha$ and two-photon emission is then expected when cooling is sufficiently strong \citep[e.g.][]{Dijkstra2009}. This process can boost the nebular contribution, particularly Ly$\alpha$ and two-photon emission, which has the effect of decreasing the overall EWs of UV emission lines, reducing the perceived ionizing photon production efficiency, and increasing the UV magnitude of a galaxy. Cooling flows have been used, for example, to explain the presence of Ly$\alpha$ blobs at intermediate redshift \citep[e.g.][]{Goerdt2010,FG2010,Rosdahl2012} and the relative contribution of cooling radiation to the total Ly$\alpha$ (and also two-photon) emission is predicted to increase substantially at higher redshift \citep[e.g.][]{Dayal2010,Yajima2012}. Given the fact that catastrophic cooling is required to drive a starburst it is reasonable to expect that cooling radiation occurs simultaneously with star formation, although the fractional contribution of cooling radiation is likely maximized immediately preceding or just after the starburst \citep[e.g.][]{Smith2019}. The exact details of how much cooling radiation can contribute to the total nebular continuum warrants further investigation under the conditions likely present in early galaxies and under conditions that may increase the electron fraction in the cooling gas (e.g. X-ray radiation). 

\paragraph{High Density Gas}
While we have primarily focused on the physical effects that can increase the nebular contribution to the UV luminosity, we also highlight the fact that certain nebular conditions can inhibit two-photon emission. For example, at high gas densities, $l$-changing collisions can shift electrons between the $2s$ to $2p$ states. However, because the Einstein coefficient is orders of magnitude larger for Ly$\alpha$ compared to two-photon emission, at high densities, Ly$\alpha$ is enhanced at the expense of two-photon emission. This is demonstrated in the bottom panel of Figure~\ref{fig:imf_temp_fneb} where we show the fractional contribution of nebular emission at 1500~\AA\ at four densities from $10^2-10^5\ {\rm cm^{-3}}$. Between $10^3\ {\rm cm^{-3}}$ and $10^4\ {\rm cm^{-3}}$ the nebular contribution begins to decrease and at a density of $10^5\ {\rm cm^{-3}}$ the decrease is much more significant. Nevertheless, at high stellar temperatures, even at extreme densities of $10^5\ {\rm cm^{-3}}$, which have only been observed in a select few high-redshift galaxies \citep[e.g.][]{Topping2024,Topping2024-NC,Senchyna2024}, the nebular continuum can still increase the UV luminosity of the galaxy by a factor of two. The electron densities of typical galaxies in the reionization epoch are such that collisional de-excitation is not expected to be very important \citep[e.g.][]{Isobe2023}.

\subsection{Observational implications}
It is clear that during a starburst with ages $\lesssim5$~Myr, the nebular continuum can represent a significant fraction of the observed SED of a galaxy. Under specific circumstances, e.g. the presence of very hot massive stars, the nebular continuum can even dominate over the stellar emission at all wavelengths observable by NIRSpec on JWST. Here we explore the observational implications for galaxies with strong nebular continuum emission.  

\subsubsection{UV luminosities and implications for the bright galaxy problem}
\label{sec:cavs_and_such}
As we have shown in Section~\ref{sec:standard_sps}, in some of the SPS models, e.g. the {\small C\&B19} models with maximum stellar masses of 300~M$_{\odot}$ and 600~M$_{\odot}$, the nebular contribution at 1500~\AA\ can reach as high as 50\%. This corresponds to a UV luminosity increase of 0.75~magnitudes. Even in the optimistic scenario of the C\&B19 models, the excess UV luminosity cannot explain the full UV variability needed to match the $z>10$ UV luminosity functions as estimated by \cite{Kravtsov2024}. However, the inclusion of the nebular emission can bring down the implied ``burstiness'' of star formation. 

\begin{figure}
\centering
\includegraphics[width=0.45\textwidth]{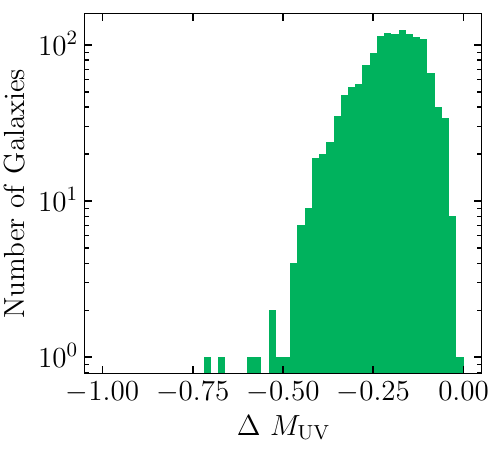}
\caption{Histogram of the change in 1500~\AA\ UV magnitude of galaxies from the {\small SPHINX} simulation when accounting for the nebular contribution to the SED. In all cases, the nebular continuum increases the UV luminosity for galaxies with realistic star formation histories, in some extreme examples by more than 0.5~magnitudes.}
\label{fig:sphinx_dm}
\end{figure}

One important caveat is that our models assume an instantaneous burst of star formation rather than a realistic star formation history. Older stars can contribute to the UV luminosity while providing no additional ionizing photons that excite the nebular continuum. The general impact of including a realistic star formation history would be to decrease the fractional contribution of the nebular continuum in the UV. To better understand this effect, we perform a simple experiment and apply the {\small C\&B19} model with a maximum mass of 600~M$_{\odot}$ (our most optimistic model) to the realistic star formation histories of simulated galaxies from the {\small SPHINX} public data release \citep{spdrv1}. The stellar masses of these galaxies range between $\gtrsim10^7\ {\rm M_{\odot}}$ to $\lesssim10^{10}\ {\rm M_{\odot}}$, and all have 10~Myr-averaged SFRs $>0.3\ {\rm M_{\odot}\ yr^{-1}}$. We then account for the nebular contribution from the young stellar populations to the total SED following our fiducial photoionization model. Figure~\ref{fig:sphinx_dm} shows a histogram of the change in UV magnitude from including the nebular continuum for $>1,400$ star-forming galaxies from the {\small SPHINX} simulation between $4.6\leq z\leq10$. While the typical magnitude increase is only $\sim0.2$ magnitudes, galaxies undergoing particularly strong bursts of star formation see much more substantial increases ($>0.5$~magnitudes) in UV luminosity due to the inclusion of the nebular continuum. Hence, during the burst, the UV magnitude is enhanced both due to the typical magnitude of the galaxy due to both the increase in SFR and the gaseous emission. We emphasize that the effect of the nebular continuum is unavoidable unless the gas is at extreme densities or the escape fraction is very high.

\begin{figure}
\centering
\includegraphics[width=0.45\textwidth,trim={0cm 0cm 0cm 1.3cm},clip]{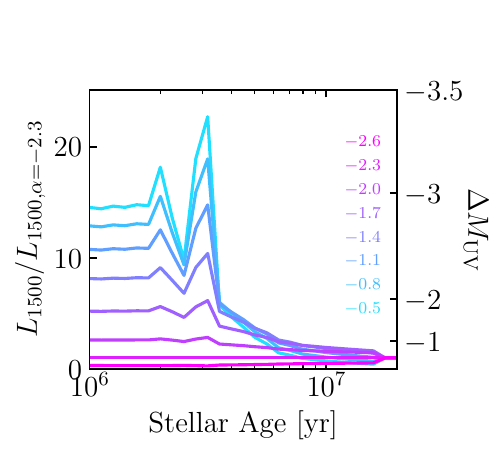}
\includegraphics[width=0.45\textwidth,trim={0cm 0cm 0cm 1.3cm},clip]{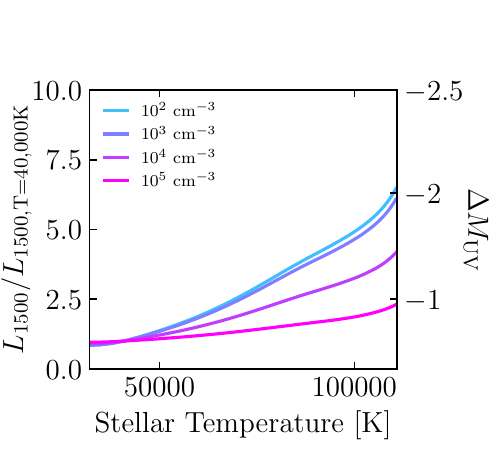}
\caption{(Top) 1500~\AA\ UV luminosity as a function of stellar age for Starburst99 models using a density of $10^3\ {\rm cm^{-3}}$ with different upper-mass IMF slopes, normalized to the model with a slope of $-2.3$. (Bottom) 1500~\AA\ UV luminosity as a function of stellar temperature for zero-age main-sequence metal-poor massive star models, normalized to a star with a surface temperature of 40,000~K. All models include both the stellar and nebular continuum. In the case of IMF variations, the number of high mass stars drives the UV luminosity increase but in the case of stellar temperature, the nebular continuum is the primary factor that leads to an increase in UV luminosity.}
\label{fig:lum_enh}
\end{figure}

Modifying the upper IMF slope of the maximum stellar mass of the IMF can similarly increase the UV luminosity of a galaxy. We demonstrate this in Figure~\ref{fig:lum_enh}. In the top panel, we show the 1500~\AA\ luminosity of the {\small Starburst99} models with different upper-mass IMF slopes as a function of stellar age, normalized by the model with an upper-mass slope of $-2.3$. A modest flattening of the slope from $-2.3$ to $-1.7$ increases the UV luminosity by a factor of five (1.75 magnitudes) at young stellar ages. Recall that under such circumstances, the nebular continuum contribution was not substantially increased and thus it is clearly the modification of the stellar component that leads to such a substantial increase in UV luminosity.

In the bottom panel of Figure~\ref{fig:lum_enh} we show the ratio of 1500~\AA\ luminosity of stars of different temperatures normalized to a 40,000~K star. Here we find that hotter stars tend to increase the UV luminosity; however, in contrast to the IMF, it is the nebular continuum that drives the substantial increase. At high gas densities, where two-photon emission is strongly suppressed, a 100,000~K star is not even twice as bright as a 40,000~K star. However, at low gas densities, when the nebular continuum is strong in the UV, the brightness can increase by a factor of five. Hence, we emphasize that there is a clear difference between simply assuming a more top-heavy IMF by modifying the upper-mass power-law slope and changing the maximum mass star in the IMF as the physical effects that lead to a luminosity increase are different. Nevertheless, the combination of both a top-heavy IMF and more massive stars can lead to substantial changes in the UV luminosity of early galaxies, especially when the nebular continuum is strong.

\begin{figure}
\centering
\includegraphics[width=0.45\textwidth]{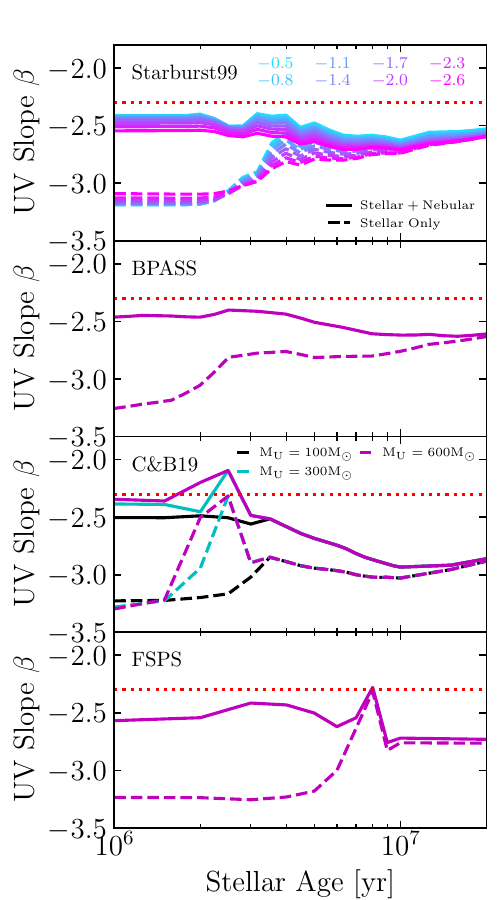}
\includegraphics[width=0.45\textwidth]{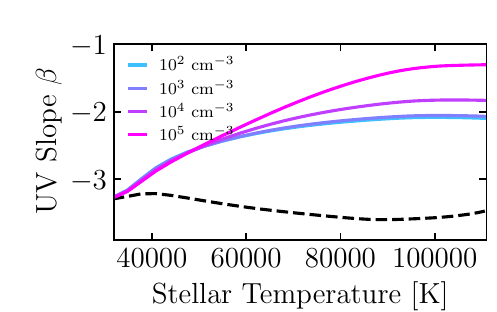}
\caption{UV slope as a function of stellar age for various SPS models (top) or stellar temperature for the individual metal-poor star models (bottom). Solid lines represent the UV slope as measured from total (stellar$+$nebular) emission while dashed lines represent the UV slope of only the stellar component. The dotted red lines indicate a slope of $-2.3$ for reference in all panels. For the Starburst99 models, colors represent different upper-mass IMF slopes (see legend in top panel), while for the C\&B19 models, colors indicate different upper-mass limits (see legend in the third panel). For the individual massive star models, colors indicate the gas density of the nebula. }
\label{fig:sps_beta}
\end{figure}

\vspace{10mm}
\subsubsection{UV continuum slopes, $\beta$}
The most common discussion on the implications of a strong nebular continuum is the impact on UV continuum slopes \citep[e.g.][]{Schaerer2002,Bouwens2010,Dunlop2013,Cullen2024,Topping2024_beta}. In particular, the nebular continuum acts to redden the observed spectrum. This is demonstrated in Figure~\ref{fig:sps_beta} where we show the intrinsic (i.e. stellar only) and observed (stellar~$+$~nebular) UV slopes for all SPS models. Here we have measured UV slopes in the wavelength range 1400~\AA\ to 2600~\AA. For young stellar ages, the intrinsic UV slopes are all bluer than $-3.0$. After the nebular continuum is added to the spectrum, the observed slope reddens to $\sim-2.5$. This is consistent with other works that have performed similar calculations \citep[e.g.][]{Schaerer2002,Bouwens2010,Dunlop2013,Cullen2024,Topping2024_beta}. 

Making the IMF more top-heavy only marginally changes the UV slope. The top panel of Figure~\ref{fig:sps_beta} shows that as the IMF upper mass slope increases from $-2.6$ to the extreme value of $-0.5$, $\beta$ only increases by 0.15 at ages prior to the main sequence lifetime of massive stars. The reason the change is so mild is that the massive stars already dominate both the UV luminosity and the ionizing photon production. Adding more massive stars increases the contribution of these stars to the 1500~\AA\ luminosity, but this effect quickly saturates and the intrinsic $\beta$ slope only becomes slightly bluer. At the same time, the nebular contribution is rising faster which reddens the slopes. Hence the net effect for a top-heavy IMF is that the UV continuum slope is redder than in the scenario with a more standard IMF (see also discussion in \citealt{Cullen2024}). We emphasize that these results are relatively independent of the chosen SPS model. Although the exact value of $\beta$ can change slightly, the trends with IMF will be similar.

To make $\beta$ even redder than in models with a top-heavy IMF, hotter stars are needed. This is shown in the bottom panel of Figure~\ref{fig:sps_beta} where UV slope is plotted as a function of stellar temperature. For typical ISM densities of $10^2-10^3\ {\rm cm^{-3}}$, $\beta$ saturates at a slope of $\sim-2.0$ consistent with earlier work that assumed Pop.~III SEDs \citep[e.g.][]{Trussler2023}. By increasing gas density, two-photon emission is suppressed and the free-bound emission dominates, which leads to extremely red UV slopes approaching $-1$. This is extremely important because we do not yet know the ISM conditions around Pop.~III stars and selection criteria should not necessarily remove redder objects as the ISM could simply be at high density. Hence we encourage photometric selection criteria to allow for these exotic scenarios.

One important caveat to these calculations is that we have computed the UV slopes without dust attenuation. Many high-redshift galaxies have Balmer decrements and UV slopes that are consistent with a dust-free scenario \citep[e.g.][]{Cullen2024,Sandles2023} and dust will only act to redden the SEDs. High quality spectra will be required to differentiate reddening due to dust versus the nebular continuum; however, it should be noted that $\beta$ is not constant with wavelength when the nebular continuum is strong. 

The modification of $\beta$ due to the nebular continuum has important implications for interpreting spectra. For example, \cite{Chisholm2022} recently proposed using the UV slope as an indicator for LyC escape. In their sample of galaxies, $\beta$ is largely set by dust attenuation and the galaxies with less dust are more likely to be LyC leakers. Applying their model to our dust-free {\small CLOUDY} simulations where $\beta$ can be as low at $\sim-2.55$, one would predict an escape fraction of $\sim17\%$ at $\beta=-2.55$. Therefore, in the zero-dust limit, this relation should be revised or not applied (see also \citealt{Choustikov2024}) while at higher metallicities, when dust is present, the relation likely remains applicable.

Another key result is that under normal nebular conditions and for typical SPS models, the UV slope should be bluer than approximately $-2.3$ for star-bursting galaxies\footnote{We emphasize that here we are discussing star-bursting galaxies. For older stellar populations, the UV slope will naturally redden. Based on the realistic star formation histories in the SPHINX simulation \citep{spdrv1}, even for galaxies with mass-weighted ages of 100~Myr, the UV slope measured from stellar$+$nebular emission remains $\sim-2.5$.}. This is indicated as the dotted red line on each panel of Figure~\ref{fig:sps_beta} and represents the maximum slope that one obtains for commonly used SPS models. Redder slopes imply either dust or an increased nebular fraction (e.g. due to hotter/more massive stars), the former of which can be tested using the ratio of Balmer lines.

\subsubsection{Ionizing photon production efficiencies, $\xi_{\rm ion}$}
The ionizing photon production efficiency $\xi_{\rm ion}$, is defined as 
\begin{equation}
    \xi_{\rm ion}=Q/L_{\nu,1500{\rm \AA}}\ {\rm erg^{-1}\ Hz},
\end{equation}
where $Q$ is the number of ionizing photons produced per second and $L_{\nu,1500{\rm \AA}}$ is the 1500~\AA\ monochromatic luminosity. Note that throughout this work, we often refer to $\log_{10}(\xi_{\rm ion})$. As $Q$ cannot be directly measured observationally, it is often inferred from emission lines (e.g. Balmer lines) using a conversion factor based on recombination theory \citep[see Appendix~\ref{app:ff} or e.g.][]{Osterbrock2006}. 

Because the nebular continuum can significantly impact the UV luminosity at 1500~\AA, measurements of $\xi_{\rm ion}$, which often normalize a Balmer line (e.g. H$\alpha$) to the 1500~\AA\ luminosity can also be impacted. More specifically, the ionizing photon production efficiency is intended to be a measurement that traces the intrinsic properties of the underlying stellar (or more generally photoionizing) population. However, if the nebular continuum represents a large fraction of the UV luminosity, these measurements will be systematically biased low. 

\begin{figure*}
    \centering
    \includegraphics[width=0.45\textwidth]{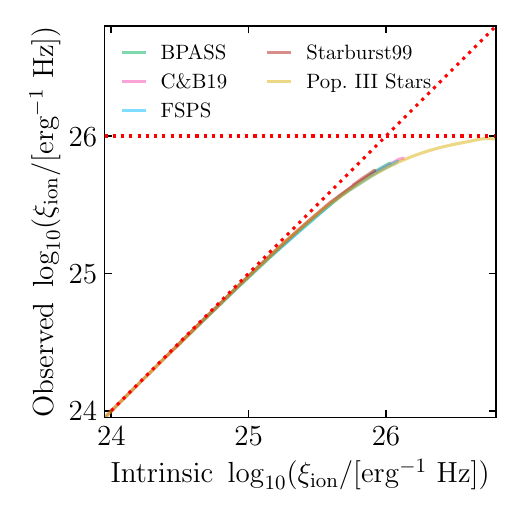}
    \includegraphics[width=0.45\textwidth]{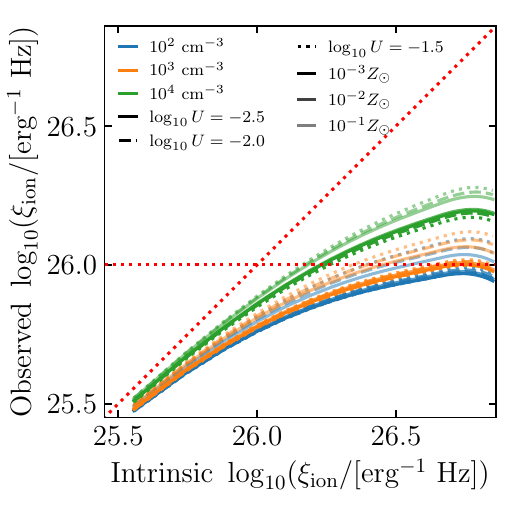}
    \caption{(Left) Observed $\log_{10}(\xi_{\rm ion})$ as a function of intrinsic $\log_{10}(\xi_{\rm ion})$ for various SPS models for our fiducial photoionization model parameters. All models follow the same trend despite differences in the underlying spectra. (Right) Observed $\log_{10}(\xi_{\rm ion})$ as a function of intrinsic $\log_{10}(\xi_{\rm ion})$ for metal-poor, hot star models for various assumptions of gas density (colors), ionization parameters (line styles), and metallicities (line opacities). This demonstrates that the properties of the nebula dictate the observed $\xi_{\rm ion}$. Note that each intrinsic $\xi_{\rm ion}$ value in the right panel corresponds to a single stellar temperature for the massive star models while in the left panel, the correspondence is with stellar age. In both panels, the diagonal dotted red line shows the one-to-one relation while the horizontal line represents $\log_{10}(\xi_{\rm ion})=26.0$.  }
    \label{fig:xi_ion}
\end{figure*}

This is demonstrated in the left panel of Figure~\ref{fig:xi_ion}. For each SPS model, we adopt the fiducial {\small CLOUDY} model and measure the intrinsic $\xi_{\rm ion}$ using the number of emitted ionizing photons ($Q$) by the intrinsic stellar continuum at 1500~\AA\ as well as the observed $\xi_{\rm ion}$ where we add the nebular continuum to $L_{\rm UV}$. Independent of the chosen SPS model, the trend is nearly identical. For $\log_{10}(\xi_{\rm ion}/{\rm erg^{-1}\ Hz})\lesssim25.5$ the intrinsic and observed $\xi_{\rm ion}$ agree. This is a result of the fact that when $\xi_{\rm ion}$ is low, there are not enough ionizing photons to create a significant nebular continuum. However, at $\log_{10}(\xi_{\rm ion}/{\rm erg^{-1}\ Hz})\gtrsim25.5$ the observed $\xi_{\rm ion}$ deviates low compared to the intrinsic value. For the highest $\xi_{\rm ion}$ values among all of our SPS models, this deviation can be nearly 1~dex. The primary conclusion here is that, while the different SPS models vary in terms of maximum intrinsic $\xi_{\rm ion}$, at high ionizing photon production efficiencies, the $\xi_{\rm ion}$ value one would measure deviates from the true value in the exact same way in all cases.

The reason why the underlying SPS models have very little impact on the relation between intrinsic and observed $\xi_{\rm ion}$ is because the contribution of the nebular continuum to the 1500~\AA\ UV luminosity is primarily sensitive to the properties of the nebula. In the right panel of Figure~\ref{fig:xi_ion} we adopt the hot star models of \cite{Larkin2023}\footnote{Although we strongly emphasize that our conclusions are independent of the chosen SPS model as we have shown. These models are adopted as they exhibit the largest range of intrinsic $\xi_{\rm ion}$.} and vary the details of the {\small CLOUDY} models such that gas density is sampled in the range $10^2-10^4\ {\rm cm^{-3}}$, the log of the ionization parameter is varied from between $-2.5$ and $-1.5$, while metallicity ranges between $10^{-3}-10^{-1}Z_{\odot}$, all of which are conditions that are thought to be generally representative of the high-redshift Universe. The models show almost no sensitivity to ionization parameter (compare line styles), while there is a weak dependence on metallicity (compare line opacities). The primary impact of metallicity is that at lower $Z$, gas temperature is higher. At higher gas temperature, two-photon emission becomes stronger partially due to the departures from Case~B \citep{Raiter2010}. This effect is particularly noticeable when the metallicity of the nebula drops to $10^{-3}Z_{\odot}$. Non-solar abundance patterns may modify where this effect matters, particularly if carbon or other strong coolants are removed from the gas. 

The largest impact on the $\xi_{\rm ion}$ discrepancy arises with gas density (compare line colors). As already explained above, this is due to $l$-changing collisions being able to shift electrons between the $2s$ and $2p$ states which then preferentially decay as Ly$\alpha$ due to the large difference in Einstein coefficients. Nevertheless, in all cases, when $\log_{10}(\xi_{\rm ion})\gtrsim25.5$, the true intrinsic value must be higher that what is observed. 

We conclude from this experiment that under typical ISM conditions, if the gas in galaxies is being irradiated by a normal stellar population, $\log_{10}(\xi_{\rm ion})$ should not be observed to be $\gtrsim25.8$. This is because the intrinsic $\log_{10}(\xi_{\rm ion})$ of even the optimistic SPS models peaks at $\sim26.1$ which corresponds to an observed $\log_{10}(\xi_{\rm ion})$ of $25.8$. Measurements higher than this value would require any one of the following:
\begin{itemize}
    \item An incorrect dust correction. For dusty galaxies UV luminosity is suppressed more than the Balmer lines which would lead to an over-prediction of $\xi_{\rm ion}$ in the case where dust acts as a screen.
    \item High gas densities such that the two-photon emission is suppressed and the true $\xi_{\rm ion}$ can be measured. 
    \item An exotic stellar population that intrinsically has a much higher $\xi_{\rm ion}$ than that predicted by normal SPS models.
    \item Extreme nebular conditions (e.g. incredibly high temperatures) or geometries, or non-stellar ionizing sources.
\end{itemize}
We discuss these options further below in the context of JWST galaxies. 

Finally, we note that the bias in $\xi_{\rm ion}$ measurements when the nebular continuum is strong is unlikely to impact inferences on reionization history. This is because in such calculations, $\xi_{\rm ion}$ is multiplied by UV luminosity density \citep[e.g.][]{Robertson2013} and even though $\xi_{\rm ion}$ is biased low, this is compensated by the fact that observed UV luminosity density is increased by the nebular continuum. 

\subsubsection{UV turnovers masquerading as DLAs}
A further implication of a strong nebular continuum is that as the ratio of the two-photon emission to the stellar continuum increases, a downturn in the UV emission just redward of Ly$\alpha$ begins to appear (see e.g. \citealt{Cameron2023_BJ}). This downturn may be perceived to be a damped Ly$\alpha$ system (DLA). The deficit in UV luminosity near Ly$\alpha$ would raise its EW; however, at high-redshift, an increasingly neutral IGM may absorb emission close to rest-frame 1216~\AA. For this reason, high-redshift galaxies with significant two-photon emission may be perceived to have strong absorption from neutral hydrogen (i.e. a DLA). 

\begin{figure}
    \centering
    \includegraphics[width=0.45\textwidth]{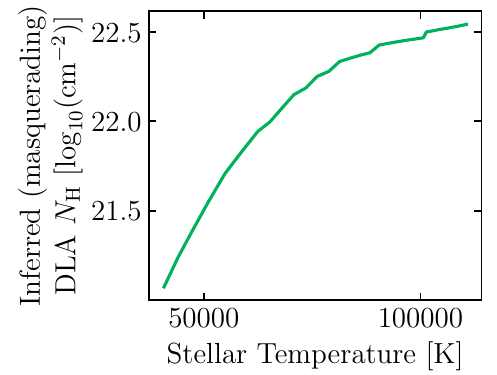}
    \caption{Strength of the UV downturn in terms of the inferred (masquerading) DLA column density as a function of stellar temperature. The UV downturn strength increases with stellar temperature as the two-photon emission becomes more visible.}
    \label{fig:dla_infer}
\end{figure}

To demonstrate this, we fit the hot star SPS models in the wavelength interval (1250~\AA~$-$~2000~\AA) using a power law~$+$~DLA model. For this simple experiment, we assume that the galaxy is at $z=9$, account for IGM absorption using the transmission curves from \cite{Garel2021}, as well as the JWST/NIRSpec Prism/Clear line spread function\footnote{Note that our results are not particularly sensitive to these assumptions.}. In Figure~\ref{fig:dla_infer} we show the inferred DLA column density as a function of stellar temperature. A clear trend emerges such that as stellar temperature increases, so does the inferred DLA column. At $T\sim100,000$~K, the inferred DLA column is $\sim10^{22.5}\ {\rm cm^{-2}}$, which is very high compared to known DLAs \citep[e.g.][]{Tanvir2019,Hu2023}, and consistent with some of the extreme systems observed at high-redshift with JWST \citep{Heintz2023,Umeda2023}.

We emphasize two key points. First, the exact value of the inferred DLA column is highly subject to the assumed intrinsic SED. For our experiment, we have fit a power law to the continuum at wavelengths $<2000$~\AA. However, the SED is not a perfect power law, noise can modify the inferred slope, and the wavelength baseline matters for the inferred UV slope. Flatter slopes lead to lower inferred column densities, while steeper slopes require more absorption. The exact shape of the underlying SED also contributes to the inferred DLA properties. Hence Figure~\ref{fig:dla_infer} should be taken more qualitatively than quantitatively (although in all cases the inferred DLA column will still be exceptionally high for high stellar temperatures). Second, the shape of a spectrum modulated by a DLA differs from the characteristic profile of two-photon emission. The DLA fits are thus not perfect representations of the true spectrum. This effect is likely diminished when noise is present and is subject to the redshift of the object as the spectral resolution of NIRSpec changes with wavelength. It should also be emphasized that strong nebular emission is not mutually exclusive with the presence of a DLA and both may occur simultaneously in the same galaxy. Nevertheless, a clear prediction of a nebula being irradiated by hot massive stars is a steep UV downturn that may masquerade as an extreme DLA. 

\begin{figure*}
    \centering
    \includegraphics[width=0.45\textwidth]{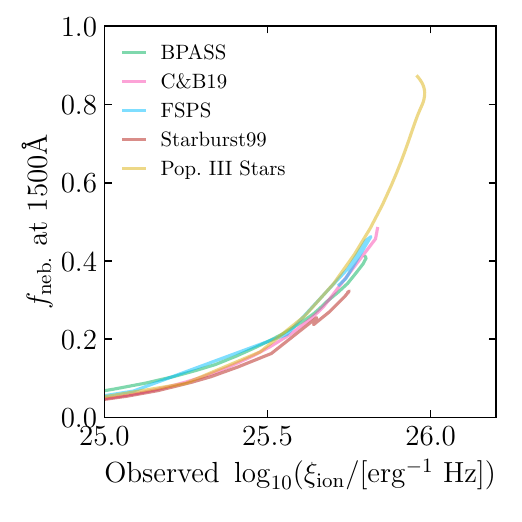}
    \includegraphics[width=0.45\textwidth]{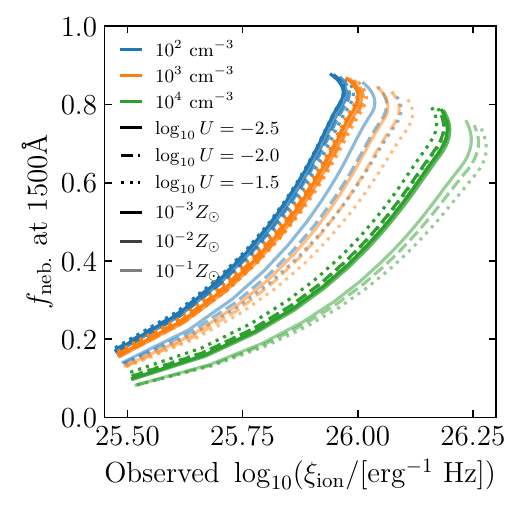}
    \caption{(Left) 1500~\AA\ nebular fraction as a function of observed $\log_{10}(\xi_{\rm ion})$ for various SPS models for our fiducial photoionization model parameters. All models follow the same trend despite differences in the underlying spectra. (Right) Observed $\log_{10}(\xi_{\rm ion})$ as a function of intrinsic $\log_{10}(\xi_{\rm ion})$ for metal-poor, hot star models for various assumptions of gas density (colors), ionization parameters (line styles), and metallicities (line opacities).}
    \label{fig:xi_ion_nf}
\end{figure*}

\subsection{Inferring a high nebular contribution}
As we have shown in Figure~\ref{fig:sps_fneb}, for typical SPS models, the nebular continuum contribution does not exceed 40\% unless more massive/hotter stars are included. Here we summarize the key characteristics of a galaxy that may have a high nebular contribution.
\begin{enumerate}
\item The presence of a Balmer jump. The only way to avoid a Balmer jump in systems with strong nebular continuum is to increase the temperature to extreme values, well beyond what are typically associated with H~{\small II} regions. In this case, the strength of the jump becomes much smaller with respect to line emission and can be dominated by other continuum processes that may make detection/identification difficult. 
\item A UV slope redder than $\sim-2.7$. The nebular continuum reddens the intrinsic stellar population and this reddening is dominated by the free-bound emission. In the extreme scenario where the continuum is dominated by two-photon emission, a lower limit of $\beta\sim-2.7$ will be measured\footnote{Note that this value is slightly sensitive to how the wavelength range over which $\beta$ is measured is set. }. Assuming the galaxy is undergoing a starburst, UV slopes redder than $\sim-2.3$ may require more extreme ionizing sources than those typically assumed in commonly used SPS models or dust attenuation.
\item A high $\xi_{\rm ion}$ as measured via a Balmer line. If $\xi_{\rm ion}$ is measured to be high, ionizing photons are both being produced at a high rate and being efficiently absorbed by the gas (i.e. not leaking such that $f_{\rm esc}\sim0$). $\log_{10}(\xi_{\rm ion})$ values above $\sim25.7-25.8$ may require more extreme ionizing sources than those typically assumed by commonly used SPS models.
\item Strong UV downturns. The appearance of a UV downturn, which can easily be confused with a high column density DLA, scales very strongly with nebular contribution due to two-photon emission. When a UV downturn appears simultaneously with bright Ly$\alpha$ emission, this is an even cleaner signature of strong nebular emission because DLAs are optically thick to Ly$\alpha$. The presence of Ly$\alpha$ is however not a necessary characteristic as both the IGM, dust, and geometry can absorb Ly$\alpha$ even if the nebular continuum is strong.  
\item High H$\alpha$ (or H$\beta$) EWs. For instantaneous bursts of star formation, the H$\alpha$ EW can reach as high as $\sim3,200$~\AA\ across all of the different SPS models. Higher values require more extreme ionizing sources. We note that high H$\alpha$ EW is not required for extreme nebular emission as there are numerous scenarios where it can be lower than predicted for the instantaneous burst. For example, the presence of an older stellar population can impact the rest frame optical while having minimal contribution in the UV. Moreover, if the two-photon emission is particularly strong, the H$\alpha$ and H$\beta$ EW would also be reduced compared to predictions from both nebular-only models that assume Case~B recombination or SPS models where the temperature is set by the SPS model.
\end{enumerate}
Any combination of the above spectral characteristics may indicate a galaxy is a candidate for having strong nebular continuum emission.

More specifically, one can even quantify the exact nebular contribution if certain properties of the gas are known. Because nebular emission is only weakly sensitive to the shape of the input ionizing spectra and strongly sensitive to the properties of the nebular gas, if the underlying physical properties of the nebula are known, the nebular contribution at any wavelength\footnote{By any wavelength we mean any wavelength where the recent star formation event dominates the spectrum.} can be uniquely determined. Hence, all one needs to infer the nebular fraction is the observed $\xi_{\rm ion}$ and a measure of gas density and temperature. We demonstrate this in the left panel of Figure~\ref{fig:xi_ion_nf} where we show the nebular contribution at 1500~\AA~as a function of observed $\xi_{\rm ion}$ (i.e. that which can be measured from the spectrum) for a metallicity of 1\%~$Z_{\odot}$ and a gas density of 100~cm$^{-3}$. The shape and normalization of the trend is relatively independent of the SPS model. The largest spread occurs between {\small FSPS} and {\small Starburst99} where at an observed $\log_{10}(\xi_{\rm ion})$ of 25.71, the nebular contribution to the total spectrum differs by 6\%. 

In the right panel of Figure~\ref{fig:xi_ion_nf} we show the 1500~\AA\ nebular fraction as a function of observed $\xi_{\rm ion}$ for a selection of the metal-poor massive star models while varying gas density, ionization parameter, and metallicity. At typical ISM densities below $10^3\ {\rm cm^{-3}}$ the nebular fraction is only sensitive to metallicity (via gas temperature). Above this critical density, gas density becomes the dominant factor as discussed in the previous section. For all gas conditions, ionization parameter is a subdominant factor.

We emphasize that there are important caveats to this analysis. First, the gas temperature in the photoionization models is determined via the gas metallicity and the underlying shape of the spectrum. Additional heating terms can modulate this relation such that lower observed $\xi_{\rm ion}$ means a higher nebular fraction. A similar behavior can be achieved by assuming non-solar abundance patterns that decrease cooling (i.e. fixing the oxygen abundance but decreasing other metal species). This is because the two-photon emission is enhanced with respect to free-bound and Balmer emission in this scenario. Likewise, any other physics, e.g. collisional excitation, cooling flows, harder radiation, that increases two-photon emission without impacting the recombination lines will mean that $f_{\rm neb.}$ is under-predicted for a given observed $\xi_{\rm ion}$.

The photoionization models also assume that the nebula is ionization bounded. In the case of density bounded nebulae, the trends can differ. Finally, there are nuances to measuring $\xi_{\rm ion}$. A fixed conversion between H$\alpha$ (or another Balmer line) and ionizing luminosity is often assumed; however, this conversion depends on temperature. Measuring the Balmer line correctly is often difficult in JWST data where the spectrum can be very noisy in the continuum near H$\alpha$. Finally, dust can significantly impact 1500~\AA\ luminosity and due to the large wavelength difference between the Balmer lines and 1500~\AA, an accurate dust correction is paramount. 

\begin{figure*}
    \centering
    \includegraphics[width=\textwidth,trim={0cm 2.15cm 0cm 3.cm},clip]{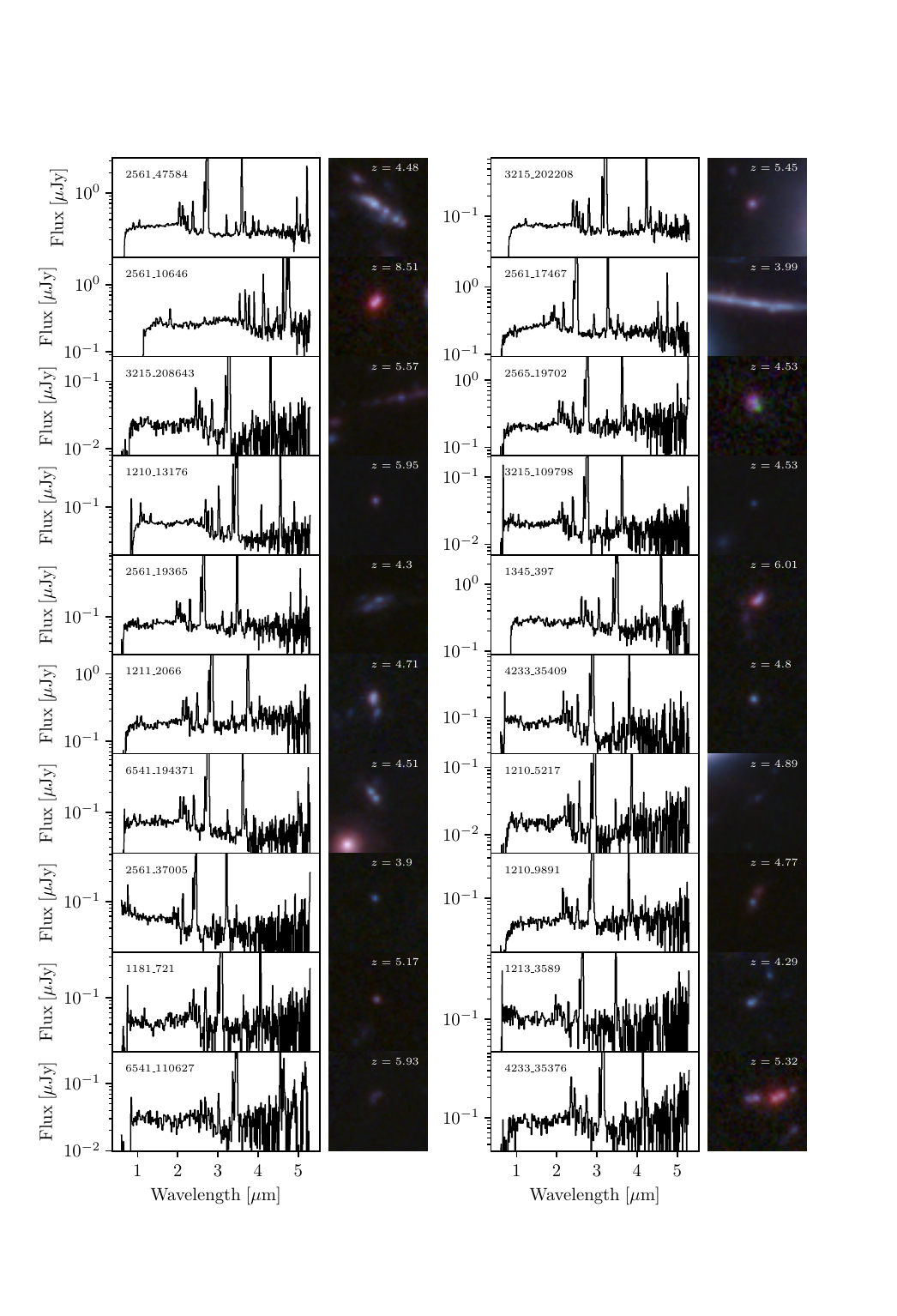}
    \caption{A gallery of example Balmer jump galaxies. We show the observed spectrum of each galaxy as well as an RGB image combining JWST NIRCam filters F444W, F200W, and F150W in the red, green, and blue channels, respectively. Each image is 1''$\times$1''.}
    \label{fig:gallery}
\end{figure*}

\subsection{Discussion}
Before proceeding to assess the contribution of nebular continuum emission to observed high-redshift galaxies, we briefly contextualize our results with respect to earlier work in the literature.

The importance of nebular emission for the SEDs of high-redshift galaxies is well established both observationally \citep[e.g.][]{Schaerer2010} and in numerical simulations \citep[e.g.][]{Wilkins2024,spdrv1}. Earlier work on Pop.~III stars \citep[e.g.][etc.]{Schaerer2002,Raiter2010,Inoue2011} were among the first to show how the nebular emission can can completely dominate over the stellar continuum in the rest-frame UV and optical under certain assumptions of the surrounding ISM (see also \citealt{Zackrisson2013,Trussler2023}), in agreement with our results here. Photoionization models with Pop.~II stars have also been analyzed in a similar context \citep[e.g.][]{Anders2003,Reines2010,Inoue2011,Byler2017} and all demonstrate that the nebular continuum is less important, albeit non-negligible, in the rest-frame UV, but can become dominant in the rest-frame optical, especially around the Balmer Jump, and at longer wavelengths. This result is consistent with our work when adopting commonly used SPS models. More quantitatively, \cite{Byler2017} showed that for stellar ages less than $3\times10^6$~Myr, the contrition of the nebular continuum to the total SED monotonically increases from rest-frame 912~\AA~to 2.3~$\mu$m. At older ages, the continuum contribution declines as the massive stars evolve off the main sequence. Here, we have only considered one gas-phase metallicity; however, \cite{Byler2017} showed that in an increasing metallicity lowers the contribution of the nebular continuum at all wavelengths. This is due to the temperature sensitivity of nebular continuum emission. Nevertheless, at the metallicites typically found in early galaxies $\lesssim0.1Z_{\odot}$ \citep{Curti2024}, the nebular contribution in the rest-frame FUV is still $\sim40\%$ according to \cite{Byler2017}, which again confirms its importance. Given the peculiar nature of early galaxies, our work expands upon these previous efforts by relaxing many of the assumptions of commonly used SPS models and highlights under what situations the nebular continuum may be more important than previously expected.

\section{A Sample of High-Redshift JWST Galaxies with Strong Nebular Continuum}
\label{sec:observations}
While the theory behind the nebular continuum is well understood, there are very few spectroscopically confirmed examples of high-redshift galaxies with a prominent nebular continuum contributions. The Balmer jump at rest-frame 3465~\AA\ is the most easily identifiable feature and this has only been reported for a select few high-redshift JWST galaxies \citep[e.g.][]{Cameron2023_BJ,Welch2024}, or in a stack of strong Ly$\alpha$ emitters \citep{Roberts-Borsani2024}. The lack of reported Balmer jumps is perhaps surprising because SED fits to NIRCam photometry often predict strong nebular continuum emission \cite[e.g.][]{Endsley2023,Bradac2024} for galaxies with extreme emission lines. For this reason, we have undertaken a survey of public JWST data to assemble a catalog of galaxies with strong nebular continuum emission.

More specifically, we visually inspected all prism spectra in v2 of the DAWN JWST Archive \citep[DJA,][]{Heintz2023} at $z\gtrsim2.5$ for a possible spectral discontinuity at the location of the Balmer jump\footnote{This possibly omits galaxies where the nebula is at such high temperatures that two-photon emission completely dominates over the Balmer jump. We leave such a search to future work.}. All spectra in this database were reduced using {\small msaexp}\footnote{\href{https://doi.org/10.5281/zenodo.7299500}{https://doi.org/10.5281/zenodo.7299500}} and full details of the reduction can be found in \cite{Heintz2023}. No other sample selection criteria were applied except for the removal of galaxies with obvious artifacts in the 2D spectra that coincidentally overlapped with the location of the Balmer jump. This means no objects were removed for obvious broad features (that may indicate the presence of an AGN) or any other reason. Our broad selection criteria leads to an extreme diversity in the types of objects that are included. Figure~\ref{fig:gallery} shows example spectra and RGB images of 20 galaxies in the data set. Our sample contains data from 14 different JWST programs as listed in Table~\ref{tab:jwstprob}, with the most coming from JADES GTO (IDs: 1180, 1181, 1210, 3215, PIs: Eisenstein, Leutzgendorf, \citealt{Bunker2023_spec,DEugenio2024}), RUBIES (ID: 4233, PI: de Graaff, \citealt{Graaff2024}), and UNCOVER (ID: 2561, PI: Labbé, \citealt{Bezanson2022}). We emphasize that because we adopt visual inspection, the sample is unlikely to be both complete and pure. The strength of spectral break that we are sensitive to is highly dependent on the signal-to-noise ratio of the continuum and we empirically find that most breaks in the sample have a $>20\%$ reduction in flux between rest-frame 3500~\AA, and 4200~\AA.

\begin{figure}
    \centering
    \includegraphics[width=0.45\textwidth,trim={0cm 0.15cm 0cm 1.0cm},clip]{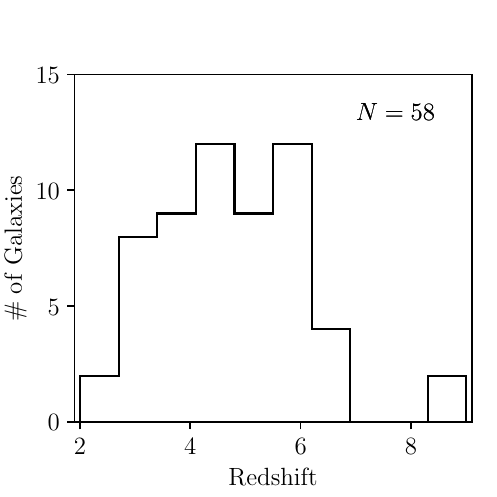}
    \includegraphics[width=0.45\textwidth,trim={0cm 0.cm 0cm 1.0cm},clip]{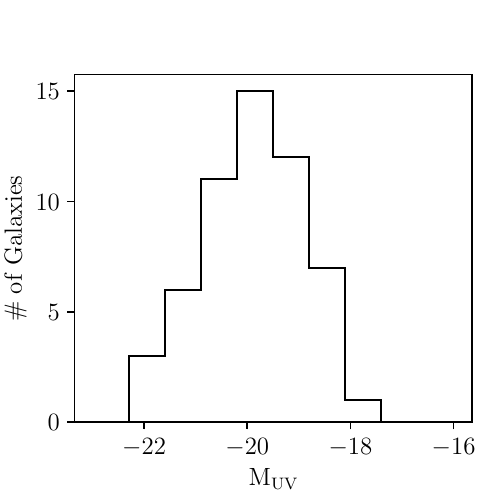}
    \includegraphics[width=0.45\textwidth,trim={0cm 0.0cm 0cm 1.0cm},clip]{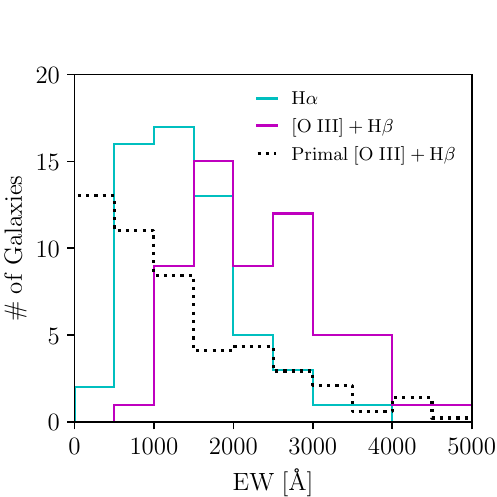}
    \caption{Histograms of redshift (top), 1500~\AA\ UV magnitude (middle), and H$\alpha$ or [O~III]$+$H$\beta$ EW (bottom) for the 58 galaxies in our dataset. For comparison, we show the [O~III]$+$H$\beta$ EW distribution from the JWST primal database \protect\citep{Heintz2024} of well-detected spectroscopically confirmed galaxies at $z>5.5$. We have renormalized the JWST primal histogram by the ratio of Balmer jump galaxies to the Primal galaxies (i.e. 58/494).}
    \label{fig:redshift_hist}
\end{figure}

The final sample contains 58 objects in the redshift interval $2.5\lesssim z\lesssim9$. Redshifts are calculated by fitting templates for emission lines to each prism spectrum or grating spectrum when available. The distribution of redshifts is shown in the top panel of Figure~\ref{fig:redshift_hist}, with most of the objects falling in the redshift interval $4\lesssim z\lesssim6$. The bias towards lower redshifts is likely a reflection of sample size and the ease of detecting the continuum at high enough signal-to-noise to identify the Balmer jump rather than strong nebular emission being more common at lower redshift. The data set spans more than four magnitudes in UV luminosity from $M_{\rm UV}>-18$ to $M_{\rm UV}<-22$ as shown in the bottom panel of Figure~\ref{fig:redshift_hist}\footnote{Note that the UV magnitudes are not corrected for magnification.}.

For each galaxy we measure various emission lines and continuum fluxes. We mask out strong emission lines and fit the remaining data with a seven parameter model using templates for the nebular continuum at various temperatures generated with {\small PyNeb} \citep{Luridiana2015} and for stellar spectra \citep{Stanway2018} including both young and old stellar populations. We use an MCMC sampler \citep{emcee2013} to measure the posterior distribution on the contributions of each component the continuum as well as the gas temperature and ages of each stellar component. To measure emission line luminosities, we randomly sample the continuum posterior and subtract it from the observed spectrum. The residuals are then fit with Gaussians using {\small specutils}\footnote{\href{https://specutils.readthedocs.io/en/stable/index.html}{https://specutils.readthedocs.io/en/stable/index.html}}, accounting for the error in the residuals. Line fluxes and their uncertainties are taken as the median and standard deviation of the Monte Carlo samples. 

Typical rest-frame equivalent widths (EWs) for H$\alpha$ in our sample of galaxies is $\gtrsim1000$~\AA. This is significantly higher than the $500-800$~\AA\ H$\alpha$ EWs found for the general population of high-redshift galaxies from \cite{Roberts-Borsani2024}. A similar trend is seen for [O~III]$+$H$\beta$ EW where the typical galaxy in our sample has an EW a few times higher than that found in \cite{Roberts-Borsani2024} and similarly that estimated for photometric samples \citep[e.g.][]{Endsley2023}. More quantitatively, comparing to the JWST Primal database \citep{Heintz2024}, the median [O~III]$+$H$\beta$ EW is 2.3$\times$ higher in our dataset compared to the typical high-redshift galaxy and the distributions are very different (see Figure~\ref{fig:redshift_hist}). Some galaxies in our sample exhibit incredibly high EWs, approaching 5000~\AA\ in [O~III]$+$H$\beta$.

\begin{table}
    \caption{Table of program name, pi, program id, and the number of galaxies from each program that are part of our Balmer jump sample. Programs are sorted descending by the number of galaxies (and then alphabetically by program name).}
    \centering
    \begin{threeparttable}
    \begin{tabular}{lllc}
    \hline
    Name & PI & ID & $N_{\rm gal}$ \\ 
    \hline
    RUBIES & de Graaff & GO 4233 & 9\\
    JADES & Eisenstein & GTO 1181 & 7\\
    UNCOVER & Labbé & GO 2561 & 7 (+1)\tnote{a} \\
    JADES & Eisenstein & GTO 1180 & 6\\
    JADES & Eisenstein & GTO 3215 & 6\\
    JADES & Luetzgendorf & GTO 1210 & 6\\
    CEERS & Finkelstein & ERS 1345 & 4\\
    DDT-EGAMI & Egami & DDT 6541 & 4 (+1)\tnote{b}\\
    DDT-CHEN & Chen & DDT 2756 & 2\\
    GO-BARRUFET & Barrufet & GO 2918 & 2\\
    WIDE & Luetzgendorf & GTO 1211 & 2\\
    DDT-COULTER & Coulter & DDT 6585 & 1\\
    GO-GLAZEBROOK & Glazebrook & GO 2565 & 1\\
    WIDE & Luetzgendorf & GTO 1213 & 1\\
    \hline
    & & Total: & 58 (+2) \\
    \hline
    \end{tabular}
    \begin{tablenotes}
    \item[a] There is an additional source in Uncover that shows a Balmer jump but v2 of the spectrum is not available in the DJA so we have excluded it from our main sample.
    \item[b] Program DDT~6541 took a spectra of the same gravitationally lensed arc that is also present in Uncover. We analyze both spectra but only count it as one system.
    \end{tablenotes}
    \end{threeparttable}
    \label{tab:jwstprob}
\end{table}

\subsection{Star Formation Histories of Balmer Jump Galaxies}
We have argued that the Balmer jump sample consists of galaxies with bursty star formation histories due to the visibility of the nebular continuum. To demonstrate this further, we have estimated the star formation history of each galaxy using SED fitting. 

We employ the {\small Dense Basis} code \citep{Iyer2017,Iyer2019}, which reconstructs the star formation history of the galaxy using Gaussian Processes. The star formation history is split into five bins of equal stellar mass. Following \cite{Iyer2019}, the underlying SPS models and nebular emission are generated with FSPS \citep{Conroy2009,Conroy2010} assuming Basel spectra and Padova isochrones. We adopt a Jeffreys prior on the shape parameters. We first construct a sparse atlas using a wide range of final stellar mass, SFR, metallicity, and $A_V$. The dust attenuation law follows \cite{Calzetti2000}. Rather than fitting the spectrum directly (as we find that many SED fitting codes struggle to simultaneously fit the Balmer jump strength and the emission line properties), we measure photometry for all JWST NIRCam wide and medium bands from the spectrum and fit the photometry. This assumes that the spectra are properly flux calibrated as we do not rescale to NIRCam fluxes apart from what was applied at the time of reduction in the DJA, i.e. no additional slit loss corrections are applied. Uncertainties on the photometry are calculated by Monte Carlo resampling the observed spectrum from the error distribution on each pixel. Because the redshift is kept fixed in the fit to that measured from the spectrum, we consider only filters that are redward of (and do not include) Ly$\alpha$. From the sparse atlas, we select the best fit model and create a refined atlas adopting priors that are $2\sigma$ deviations around the best model. Our final best model comes from the refined atlas. 

\begin{figure}
    \centering
    \includegraphics[width=0.45\textwidth,trim={0 0cm 0 0cm},clip]{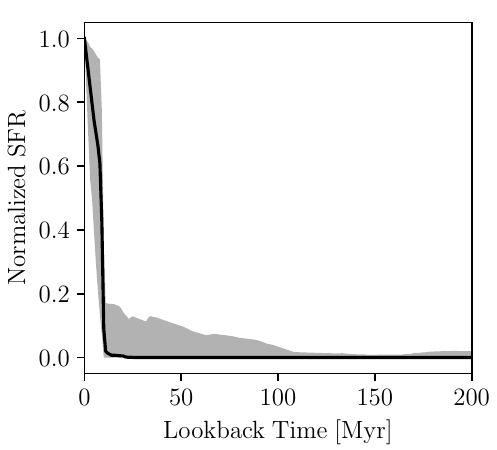}
    \caption{Normalized stacked star formation histories of the Balmer jump galaxies. The star formation history of each galaxy is normalized so that the SFR at the present time is 1. The black line and gray shaded region show the median star formation history and the $1\sigma$ scatter.}
    \label{fig:SFH}
\end{figure}

In Figure~\ref{fig:SFH}, we show the normalized star formation history stack of all of the Balmer jump galaxies over the past 200~Myr. For each galaxy, we normalize the star formation history by the SFR at the current time. The histories are then interpolated onto the same time axis and we show the median and $1\sigma$ deviations. 

The star formation histories appears to be nearly a delta function at the present time. The median history prefers almost no star formation beyond $\sim35$~Myr and the present day SFR of the median history is $>200\times$ that at $20$~Myr. This exercise clearly demonstrates that the Balmer jump sample represents a unique class of galaxies undergoing extreme bursts of star formation, which potentially makes them analogues of the extreme high-redshift galaxy population.

\begin{figure*}
    \centering
    \includegraphics[width=\textwidth]{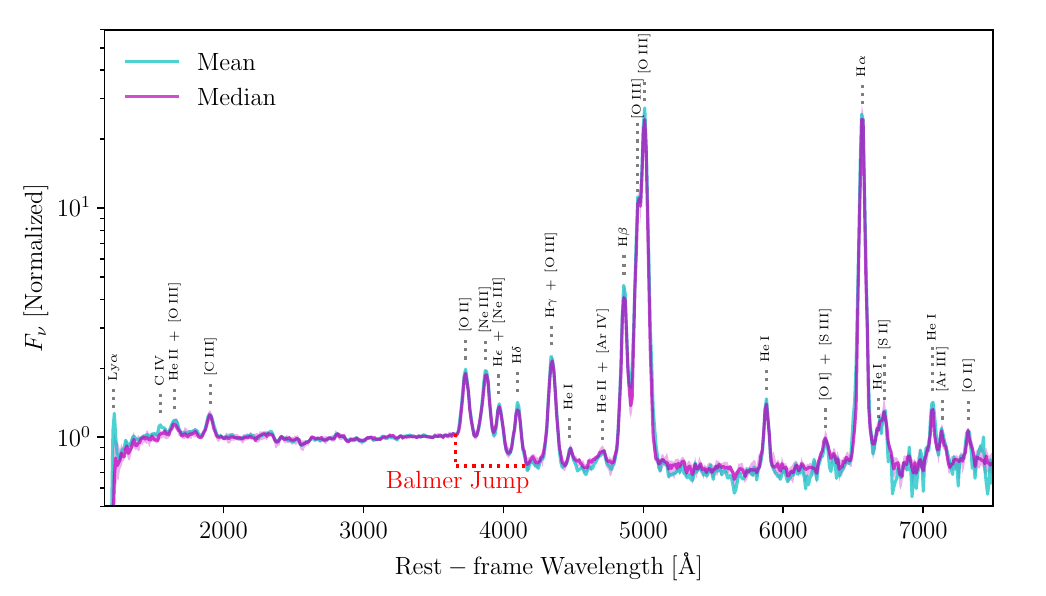}
    \caption{Mean (cyan) or median (magenta) stack of all 58 galaxies in our sample. We highlight the spectral discontinuity at the location of the Balmer jump. The shaded region around the median stack shows the 5th-95th percentile distribution calculated from 10,000 bootstrap resamples. This demonstrates that no single galaxy has an extreme impact on the stack.}
    \label{fig:stack}
\end{figure*}

\subsection{The Characteristic Spectrum of Strong Nebular Emission}
We begin by stacking the galaxies to increase the signal-to-noise. Each prism spectrum is de-redshifted to the rest-frame and we use the {\small FluxConservingResampler} from {\small specutils} to resample the spectra onto a common wavelength grid. Resampled spectra are then normalized to have a flux density of unity at 3200~\AA. Mean and median stacks are shown in Figure~\ref{fig:stack}.  Scatter in the median stack is calculated from 10,000 bootstrap resamples. Apart from exhibiting very strong emission lines, both in the rest-frame UV and optical, the most notable feature is the strong spectral discontinuity at 3645~\AA\ which demonstrates that these galaxies exhibit a very clear Balmer jump.

\begin{figure*}
    \centering
    \includegraphics[width=\textwidth]{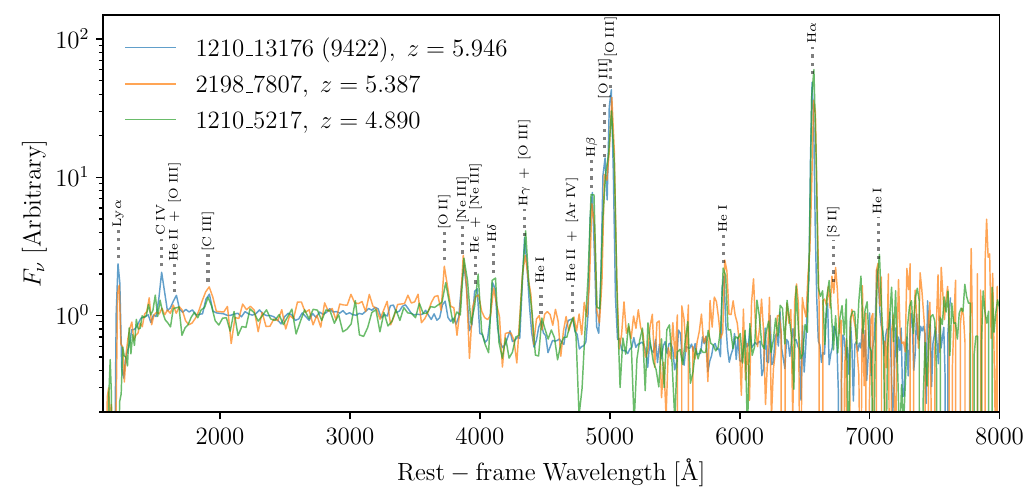}
    \caption{Comparison of 1210\_5217 at $z=4.890$ with potential nebular-dominated galaxies 2189\_7807 ($z=5.387$) and GS\_9422 ($z=5.946$). All three show a clear Balmer jump and a very similar UV downturn. Spectral modeling predicts a very high nebular fraction in each of these objects.}
    \label{fig:two_phot_spectra}
\end{figure*}

Only minor differences are apparent between the mean and median stacks. Both exhibit remarkably little variation in UV flux as a function of wavelength between 1500~\AA\ and 3000~\AA, indicating a spectral slope of $\beta\sim-2$. The strength of the Balmer jump is also similar and both stacks exhibit a downturn in the UV, just redward of Ly$\alpha$. Emission line fluxes tend to be slightly higher for the mean stack. This is particularly true for C~{\small IV}~$\lambda\lambda$1550 and Ly$\alpha$, where we find that fewer than half of the Balmer jump galaxies are strong Ly$\alpha$ emitters.

The stack exhibits strong rest-frame UV and optical emission lines, consistent with the individual EWs measured in our sample. There is also evidence for slightly rarer emission lines such as the combination of He~{\small II}~$\lambda$4686 and the nearby [Ar~{\small IV}]~$\lambda\lambda$4711,4740 doublet. The [O~{\small II}] quadruplet auroral line feature at rest-frame 7320-7330~\AA\ is clearly visible as was also recently seen in the lower-redshift stack of \cite{Strom2023}. Numerous He~{\small I} lines are present as well as weak [Ar~{\small III}] at~7136~\AA.

\subsection{The nebular contribution in Balmer jump galaxies}
Now considering individual galaxy spectra, one of the key predictions of the SPS models is that the nebular contribution at 1500~\AA\ maximizes at $\sim40\%$ unless hotter/more massive stars are included. Having a sample of galaxies with bursty star formation histories and clear nebular continuum contributions in the optical, we can test whether this $\sim40\%$ upper limit holds for real galaxies. 

Inferring the nebular contribution is not straight-forward unless the detailed properties of the nebula are known. Robust temperatures, densities, and $\xi_{\rm ion}$ values are not available for the vast majority of galaxies in our sample and thus we resort to an alternative technique. In principle, the shape of the continuum encodes the contributions of stellar and nebular emission. However, the key confounding factor occurs in the UV where downturns can occur both due to neutral hydrogen absorption and two-photon emission. We thus fit the galaxies twice, once using our seven parameter MCMC fitting code and a second time allowing for a DLA in the fit. We then focus on sources where the nebular continuum is predicted to be strong independent of the model.

\subsubsection{Nebular dominated galaxies}
Based on this procedure, a few galaxies immediately appear to require an unusually strong nebular contribution. The first galaxy is 1210\_13176 at $z=5.943$ (also referred to as NDG\_9422 in \citealt{Cameron2023_BJ}), the spectrum of which is shown as the blue line in Figure~\ref{fig:two_phot_spectra} and the spectral properties can be found in Table~\ref{tab:NDGs}. Our template fitting indicates a $1\sigma$ lower limit of 80\% for the nebular contribution to the 1500~\AA\ luminosity. This galaxy is discussed at length in \cite{Cameron2023_BJ} and the spectrum was first described in \cite{Saxena2024}, so here we only highlight the key features.

What makes 1210\_13176 interesting is that not only does it exhibit a very strong downturn in the UV, which if interpreted as a DLA would imply an extreme neutral column density of $\sim10^{23}\ {\rm cm^{-2}}$, it also has strong Ly$\alpha$ emission. If a such a high column density DLA is present, it remains unclear why such a large fraction ($>25\%$, \citealt{Saxena2024}) of the Ly$\alpha$ escapes, especially considering that the galaxy is very compact with no visible companion that could otherwise account for the Ly$\alpha$ emission. Interestingly, \cite{Cameron2023_BJ} demonstrated that the shape of the continuum is consistent with the expectations of the nebular continuum at a gas temperature similar to that measured from [O~{\small III}]~$\lambda$4363/[O~{\small III}]~$\lambda$5007. The $\log_{10}(\xi_{\rm ion}/{\rm erg^{-1}\ Hz})$ for this galaxy is 25.8, indicating that the nebular contribution in the UV could be very strong. While \cite{Cameron2023_BJ} present numerous physical mechanisms to explain the shape of the spectrum, hot stars remain a natural explanation for the continuum shape of this galaxy. 

More recently, \cite{Schaerer2024b} have argued, based on Case~B recombination, that the H$\beta$ equivalent width is too low for the spectrum to be fully nebular dominated. We argue that the discussion presented in \cite{Schaerer2024b} does not necessarily apply to this object. It is clear from the Balmer decrement that the Case~B assumption is not valid for this galaxy \citep[e.g.][]{Yanagisawa2024}. 
Moreover, the temperature inferred from both the [O~{\sc iii}] auroral line ratio and the Balmer jump is $T_e>18,000$~K, much higher than the typically assumed $10^4$~K. At these temperatures, the expected Case~B H$\beta$ EW is 860~\AA\ (somewhat lower than the $\sim$1,000~\AA\ suggested in \citealt{Schaerer2024b}, under the assumption of $T_e=15,000$~K).

More importantly, as shown in Appendix~\ref{app:spec_prob}, the two-photon emission still contributes significantly in the rest-frame optical. Collisional excitation to the $n=2$ state can significantly enhance the two-photon continuum, further lowering the observed H$\beta$ EW. This has, in fact, been predicted for hot and metal-poor environments \cite{Raiter2010,Ribas2016}. The best-fit model of \cite{Cameron2023_BJ} predicts such an enhancement which naturally explains both the observed H$\beta$ and H$\alpha$ EW. Finally this system shows clear oxygen, carbon, and neon emission, which indicates that there would have been a previous generation of stars. This would further lower the Balmer EWs without impacting the fit in the UV. Thus, as presented in \cite{Cameron2023_BJ}, we stress that the spectrum of 1210\_13176 is fully consistent with having a dominant contribution from the nebular continuum in the UV.

Other studies have suggested alternative explanations for the observed UV downturn and concomitant Ly-$\alpha$ detection. Based on the fact that several other galaxies in the same field as 1210\_13176 have DLAs, \cite{Heintz2024_nat,Terp2024} argue that a lower-redshift DLA at $z\sim5.4$ in a foreground galaxy cluster would offset the Ly$\alpha$ absorption trough, which could allow Ly$\alpha$ to leak and reconcile the differences in shape between the observed UV downturn and a DLA. However, this solution would imply an even larger gas column density. Likewise, \cite{Tacchella2024,Li2024} argue that an obscured black hole with an offset narrow line region and a large DLA column could provide a similar spectral shape. 
Because of the UV downturn and strong Ly$\alpha$, any solution to this puzzle that invokes a DLA requires a ``fine-tuned'' geometry. Therefore we argue that, for a single object these could be reasonable explanations; however, if more galaxies like 1210\_13176 are identified, such models become less favorable. 

\begin{table*}
    \caption{Properties of five galaxies predicted to have nebular contributions far exceeding that predicted from SPS models. We list the galaxy ID, redshift, 1500~\AA\ UV magnitude, UV slope, Balmer decrement, EWs of H$\alpha$ and [O III]+H$\beta$, H~II region temperature predicted from the MCMC fitting of the spectra, ionizing photon production efficiency, $1\sigma$ lower limit on the estimated nebular contribution at 1500~\AA, and finally the DLA gas column density or $1\sigma$ upper limit. For each galaxy we list two values of $\xi_{\rm ion}$ computed using either H$\alpha$ (top) or H$\beta$ (bottom). The first three galaxies may host a population of massive stars while the bottom two spectra of different regions of a gravitationally lensed arc appear to be consistent with purely Case~B recombination.}
    \centering
    \begin{tabular}{lcccccccccc}
    \hline
    ID & $z$ & M$_{\rm UV}$ & $\beta$ & H$\alpha$/H$\beta$ & H$\alpha$ EW & [O III]+H$\beta$ EW & $T_{\rm HII}$ & $\log_{10}(\xi_{\rm ion})$ & $f_{\rm neb., 1500\AA}$ & $N_{\rm H,DLA}$\\
     & & & & & [\AA] & [\AA] & [K] & $[{\rm erg^{-1}\ Hz}]$ & & $\log_{10}({\rm cm^{-2}})$ \\
    \hline
    1210\_13176 & 5.934 & $-19.65^{+0.01}_{-0.01}$ & $-2.00^{+0.02}_{-0.02}$ & $2.54^{+0.02}_{-0.03}$ & $1,871^{+26}_{-26}$ & $3,461^{+44}_{-44}$ & $20,156^{+1,161}_{-1,076}$ & $25.78^{+0.01}_{-0.01}$ & $>80\%$ & N/A \\
     & & & & & & & & $25.81^{+0.01}_{-0.01}$ & & \\
    2198\_7807   & 5.387 & $-19.80^{+0.03}_{-0.04}$ & $-1.85^{+0.08}_{-0.08}$ & $2.89^{+0.12}_{-0.13}$ & $1,527^{+83}_{-77}$ & $2,834^{+155}_{-169}$ & $24,426^{+7,227}_{-4,911}$ & $25.81^{+0.02}_{-0.03}$ & $>60\%$ &  N/A \\
     & & & & & & & & $25.78^{+0.03}_{-0.03}$ & & \\
    1210\_5217   & 4.888 & $-17.88^{+0.04}_{-0.03}$ & $-1.99^{+0.07}_{-0.09}$ & $2.97^{+0.06}_{-0.05}$ & $3,323^{+165}_{-152}$ & $3,357^{+148}_{-133}$ & $20,492^{+6,350}_{-4,746}$ & $26.04^{+0.03}_{-0.03}$ & $>63\%$ & $ <21.81$ \\
     & & & & & & & & $26.01^{+0.02}_{-0.02}$ & & \\
    \hline
    2561\_17467 & 3.990 & - & $-1.58^{+0.02}_{-0.02}$ & $2.82^{+0.04}_{-0.05}$ & $1,545^{+13}_{-13}$ & $3,257^{+63}_{-34}$ & $25,422^{+1,246}_{-1,128}$ & $25.97^{+0.01}_{-0.01}$ & $>61\%$ & $22.50^{+0.12}_{-0.17}$ \\
    & & & & & & & & $25.95^{+0.01}_{-0.01}$ & & \\
    2756\_301 & 3.990 & - & $-1.33^{+0.03}_{-0.03}$ & $2.80^{+0.13}_{-0.11}$ & $1,663^{+29}_{-30}$ & $3,429^{+89}_{-85}$ & $41,814^{+2,868}_{-2,760}$ & $26.03^{+0.01}_{-0.01}$ & $>99\%$ & $22.75^{+0.12}_{-0.17}$ \\
    & & & & & & & & $26.00^{+0.02}_{-0.02}$ & & \\
    \hline
    \end{tabular}
    \label{tab:NDGs}
\end{table*}

Interestingly, a second galaxy in our sample, 2198\_7807 at $z=5.387$ from the Cycle~1 program: `Quiescent or Dusty? Unveiling the Nature of Red Galaxies at $z > 3$' (Program ID: 2198; PIs: L. Barrufet \& P. Oesch; \citealt{Barrufet2024}) shows a very similar UV downturn and strong Ly$\alpha$ emission as 1210\_13176 (see the orange line in Figure~\ref{fig:two_phot_spectra}. This galaxy is remarkably similar to 1210\_13176 in UV magnitude, $\beta$ slope, and $\xi_{\rm ion}$, and has similarly high emission line EWs (see Table~\ref{tab:NDGs}). There is no indication of dust in the object and our $1\sigma$ lower-limit on the 1500~\AA\ nebular fraction of 60\%, much higher than that of typical SPS models (see Figure~\ref{fig:sps_fneb}). As these two spectra are so similar, we refrain from a more detailed discussion since the same arguments presented for 1210\_13176 in \cite{Cameron2023_BJ} apply to 2198\_7807. However, we briefly highlight a few key differences. The C~{\small IV}~$\lambda\lambda$1550 emission is much weaker in 2198\_7807 and there is no evidence for He~{\small II}~$\lambda$1640, which suggests the emission is even less likely to be powered by a black hole\footnote{Although we emphasize that for 1210\_13176, nebular diagnostics suggest that this object is not AGN dominated or at best a composite \citep{Cameron2023_BJ}.}. [O~{\small II}]~$\lambda\lambda$3727 is also stronger in 2198\_7807 compared to 1210\_13176, suggesting it is less likely that the emitting H~{\small II} regions are density bounded. Because 2198\_7807 is at a redshift consistent with or lower than the presumed proto-cluster that was used in \cite{Heintz2024_nat,Terp2024} to explain both the UV downturn and the Ly$\alpha$ emission in 1210\_13176, the same argument cannot apply to 2198\_7807, unless there is another foreground cluster at even lower redshift. This would be a remarkable coincidence given that the shapes of the UV spectra are so similar. Finally, fine-tuned models that invoke strong DLAs \citep[e.g.][]{Tacchella2024,Li2024} are less appealing given that there is more than one spectra with an almost identical UV shape.

A third galaxy that is predicted to have a high 1500~\AA\ nebular fraction is 1210\_5217 at $z=4.89$ (green line in Figure~\ref{fig:two_phot_spectra}). This galaxy was presented in \cite{Boyett2024} where they noted an extremely high $\log_{10}(\xi_{\rm ion})$ value of $25.9\pm0.1$ and no dust. Such a value is much higher than expected for commonly used SPS models when the nebular continuum is taken into account (see above). This value is also consistent with that measured in our work ($26.04^{+0.03}_{-0.03}$) to $\sim1\sigma$. This UV luminosity of this galaxy is much fainter by two magnitudes compared to the previous two discussed in this section, but it has a very similar UV slope of $-1.99^{+0.07}_{-0.09}$. As this is much redder than that expected from SPS models for a single burst and given the lack of dust in this object, a high UV nebular fraction is a likely explanation. Moreover, this galaxy has extreme emission line equivalent widths. 
The equivalent widths of H$\alpha$ and [O~{\small III}]~$+$~H$\beta$ are both $\gg3,000$~\AA, the highest H$\alpha$ EW among the entire Balmer jump sample. No He~{\small II}~$\lambda$1640 or $\lambda$4686 is detected in the prism or gratings, despite a clear grating detection of O~{\small III}~$\lambda$1666, and strong line ratios indicate that this source is not a candidate AGN \citep{Scholtz2023,Curti2024}.

While 1210\_5217 is not a Ly$\alpha$ emitter, it also exhibits a strong UV downturn that has a remarkably similar shape to 1210\_13176 and 2198\_7807 (Figure~\ref{fig:two_phot_spectra}). Even allowing for a DLA in our MCMC fit, the shape of the UV downturn prefers a nebular solution and hence we find a $1\sigma$ lower limit of 63\% for the nebular fraction at 1500~\AA. Like the other two galaxies, the gas temperature is predicted to be much higher than local H~{\small II} regions, which could enhance the nebular contribution in the UV. We emphasize that a lack of Ly$\alpha$ emission does not preclude 1210\_5217 from being nebular dominated. The IGM is not completely transparent at these redshifts \citep[e.g.][]{Inoue2014} and the actual emission profile of Ly$\alpha$ is highly viewing-angle dependent, even for galaxies without DLAs \citep[e.g.][]{Blaizot2023}. The presence of massive stars is a natural explanation for the UV downturn and the relatively red $\beta$ slope given the high $\xi_{\rm ion}$. If such stars are common at high redshift, we may expect more to look like 1210\_5217 than either 1210\_13176 or 2198\_7807 due to the sight-line variability of Ly$\alpha$ and the fainter UV magnitude. 

\subsubsection{No young stars in the shutter?}
One of the key features of the three galaxies in the previous section is that their Balmer decrements indicate that they are dust-free but their UV slopes are redder than that of typical SPS models. Redder slopes can be achieved if the intrinsic ionizing photon production efficiency is higher than the SPS models which enhances the nebular contribution at all wavelengths. Such is the case when hot/massive stars are present (see Section~\ref{sec:hot_massive} and also \citealt{Schaerer2024b}). However, even in the case of hot/massive stars, if the gas density is $\lesssim 10^4$ cm$^{-3}$, the UV slope for a young stellar population cannot be redder than $\beta\sim-1.7$. If a Balmer jump is present in the spectrum and the Balmer decrement indicates that there is no dust in the system, a $\beta$ redder than $\sim-1.7$ would either imply that the gas is either exceptionally dense, suppressing the two-photon continuum, or there is no underlying young stellar continuum and we are observing pure nebular emission in the UV. As we show in Appendix~\ref{app:spec_prob}, in the latter scenario, $\beta$ from pure Case~B recombination can approach values of $\sim-1.3$.

\begin{figure}
    \centering
    \includegraphics[width=0.45\textwidth]{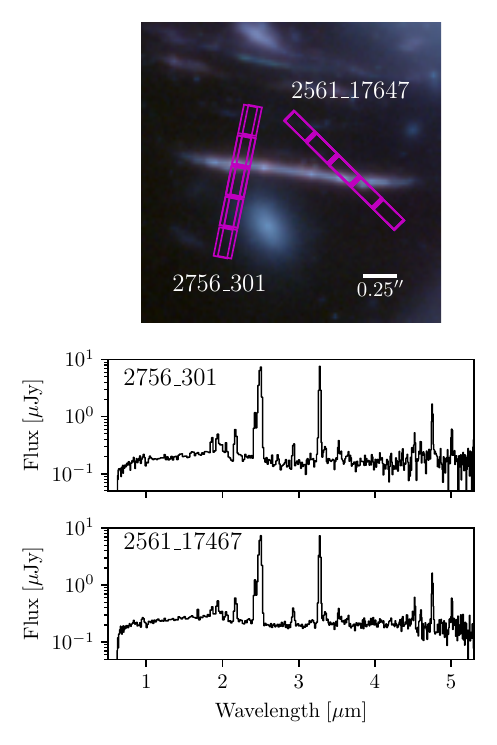}
    \caption{(Top) RGB image of the Cosmic Gummy Worm combining JWST NIRCam filters F444W, F200W, and F150W in the red, green, and blue channels, respectively. Overlaid in magenta are the locations of the MSA shutters. (Bottom) Prism spectra of two regions of the gravitationally lensed arc at $z=3.99$, both showing red continua and a Balmer jump.}
    \label{fig:crazy_lensed}
\end{figure}

The latter scenario where there is little or no young stellar population can occur if the NIRSpec slit only covers the nebula. We find two example spectra where this may be occurring. In Figure~\ref{fig:crazy_lensed} we show an image of a gravitationally lensed arc (The Cosmic Gummy Worm) being lensed by the Abell~2744 cluster. Two separate JWST programs (2561, PI: Labbé and 2756, PI: Egami) placed shutters on different parts of the arc and the spectra are shown in the bottom two panels of Figure~\ref{fig:crazy_lensed}. This source has already been discussed extensively in the literature \citep{Vanzella2022,Lin2023} and is thought to host proto-globular clusters. Most notably, \cite{Lin2023} provide a map of UV slope across the arc and find values ranging from $-2.5$ to $-1.5$ with the bluest values centered on the location of where there are thought to be young stars. Both previous analyses conclude that this system likely hosts a young, metal-poor stellar population with relatively little dust.

The key features of these spectra (see Table~\ref{tab:NDGs}) are that their Balmer decrements are consistent with Case~B values indicating that little or no dust is present in the object and the $\log_{10}(\xi_{\rm ion})$ values are very high at $25.97$ or $26.03$. The UV slopes are very red with values of $-1.58$ and $-1.33$, which is much redder than can be generated by even the hot star models, unless extreme densities are assumed. Both spectra exhibit exceptionally high [O~{\small III}]+H$\beta$ EWs $>3,000$~\AA.

What is unique about our MCMC fits to these spectra is that not only do they require a very high nebular fraction at 1,500~\AA\ that is significantly higher than what can be produced with typical SPS models, especially 2756\_301 where the model prefers a fully nebular solution\footnote{Note that the fit for 2756\_301 requires a temperature extrapolation for the nebular continuum and there is a systematic uncertainty, particularly related to temperature associated with this fit.}, but the model also simultaneously infers a very strong DLA column between $10^{22.50}-10^{22.75}\ {\rm cm^{-2}}$. These two spectra are the only ones in our sample that exhibit such behavior. The continuum shapes of these spectra are consistent with being nebular continuum from Case~B recombination with a DLA.

In fact, this system is already known to host a DLA based on shorter wavelength data from MUSE that covers Ly$\beta$ \citep{Lin2023}, with those authors inferring a DLA column of $N_{\rm H}\sim10^{21.8}~{\rm cm^{-2}}$. The DLA column likely varies across the system which may explain why our measurement, which only covers part of the arc, finds a higher column density. 

As speculated by \cite{Lin2023}, our results support a scenario in which the red portions of the arc are dominated by the nebular continuum. Spectra such as 2561\_17467 and 2756\_301 are extreme examples where gravitational lensing can help distinguish the gas from the underlying continuum source and hence the spectra may appear as being nebular dominated. More such examples exist in the DJA; however, none are as exemplary as those shown here. The high measured $\xi_{\rm ion}$ values are thus not necessarily reflective of the underlying ionizing source population, but rather a manifestation of separating a continuum source from the nebula.

It should be emphasized that this system is not a Ly$\alpha$ emitter. This is a clear example of a galaxy with a very high, spatially varying DLA column density, and yet the spectrum looks completely different from the objects above with strong UV downturns and Ly$\alpha$ emission. If the ISM of this object has any comparison to 1210\_13176 and 2198\_7807, it makes it even more peculiar that if a high column density DLA was the explanation for their spectrum, Ly$\alpha$ emission can leak significantly from those objects but not for 2561\_17467 and 2756\_301.

\section{Implications for redshift $>10$ galaxies}
\label{sec:discuss}
Both the theoretical and observational exercises presented here provide insight into the expected properties and physical nature of galaxies that are undergoing extreme bursts of star formation. Nearly all numerical simulations that attempt to resolve the ISM predict that star formation becomes increasingly bursty at high-redshift \citep[e.g.][]{Ma2018,spdrv1,Pallottini2022}. Moreover, the leading explanation for the high-redshift bright galaxy problem is that surveys are sampling the extreme ``tip-of-the-iceberg'' of bursty star formation \citep[e.g.][]{Ren2019,Mason2023,Shen2023,Sun2023,Kravtsov2024}. For both of these reasons, the nebular continuum is thus expected to become increasingly important at $z>10$. 

In this Section, we discuss some of the physical properties of known $z>10$ galaxies in the context of the models we have presented for the nebular continuum which may help elucidate their physical origin. We emphasize that in order to explain the high-redshift bright galaxy problem, only a subset of the known objects would require deviations from standard galaxy formation physics. The expectation is that many objects will be in agreement with commonly used SPS models of bursty galaxy formation, while others may require a different interpretation.

\subsection{Do the bright high-redshift galaxies have strong nebular continuum emission?}
Except for extreme examples like GNz11 where the continuum is detected and a Balmer jump detection is not conclusive\footnote{In one pointing, a spectral discontinuity appears at the location of the Balmer jump. Since not all observations show this, the presence of a Balmer jump is inconclusive. From photometry, there may be some hint that the central region is consistent with strong nebular emission \citep{Tacchella2023}.}, the signal-to-noise of the spectra of known $z>10$ galaxies at rest-frame 3,645~\AA\ is too low to see a Balmer jump. Stacking all publicly available spectra in the DJA with $z>10$ also leads to a too low signal to detect a Balmer jump. Thus we discuss these high-redshift galaxies using more indirect indicators of strong nebular emission. 

\begin{figure}
    \centering
    \includegraphics[width=0.45\textwidth]{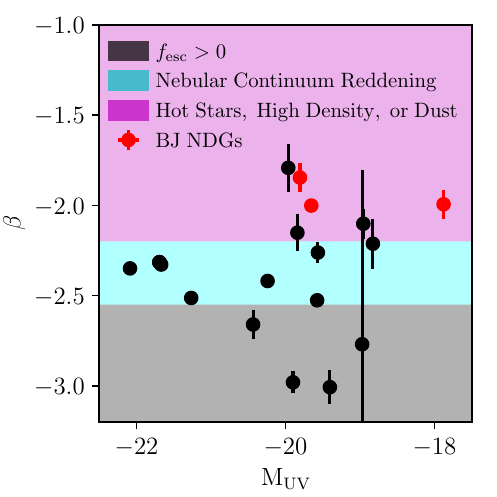}
    \caption{UV continuum slope, $\beta$, as a function of M$_{\rm UV}$ for all $z>10$ galaxies in the JWST Primal database \protect\citep{Heintz2024} are shown as black points. The region shaded in cyan show the expected UV slopes from all of the different SPS models that we test. Slopes bluer than this (shaded in black) would require a non-zero escape fraction, while redder slopes (magenta), would indicate the presence of either hotter stars, the combination of hotter stars and high gas density, or dust. Red points represent the three galaxies in our Balmer jump sample that appear to have a very strong nebular contribution at 1500~\AA. We emphasize that this diagram should only be applied to high EW galaxies where the UV slope is reddened due to nebular emission or dust, rather than an aging stellar population.}
    \label{fig:primal_beta_muv}
\end{figure}

Unfortunately, very few properties can be measured for most $z>10$ galaxies apart from their UV magnitudes, continuum slopes, and an occasional emission line. Thus in Figure~\ref{fig:primal_beta_muv}, we plot $\beta$ as a function of M$_{\rm UV}$ for all galaxies in the JWST Primal sample \citep{Heintz2024}. The region shaded in cyan show the expected UV slopes from all of the different SPS models that we test. Slopes bluer than this (shaded in black) would require a non-zero escape fraction, while redder slopes (magenta), would indicate the presence of either hotter stars, the combination of hotter stars and high gas density, or dust. 

Most of the $z>10$ galaxies fall close to the cyan region indicating that vanilla nebular continuum presents a reasonable explanation. Recall that this does not rule out the presence of a flatter upper-mass slope, as long as the upper mass does not increase much above $\sim100-300\ {\rm M_{\odot}}$. Three galaxies have $\beta<-2.55$ which would imply a weaker nebular continuum contribution. This could be caused by a non-zero escape fraction, or if the massive stars output fewer ionizing photons than expected compared to the SPS models that we consider. Finally, 3/17 galaxies have $\beta>-2.2$, although within the uncertainties, two are consistent with the cyan region. The outlier is the $z=11.5$\footnote{Note that this is the same galaxy as reported in \cite{Haro2023} with MSA ID 10 (CEERS2\_588). The Lyman break and possible detection of [O~II]~$\lambda\lambda$3727 provide different redshifts. Following the JWST primal database, we adopt the redshift from the break.} galaxy (CEERS2\_588) that has $\beta=-1.79\pm0.13$, more than $3\sigma$ away from the vanilla prediction. 

Perhaps the simplest explanation for this galaxy is that it is dusty. However, in the context of theoretical models, this explanation is unsatisfying because most pre-JWST predictions of the high-redshift bright galaxy number counts from full-box simulations catastrophically fail in reproducing observations \citep[e.g.][]{Finkelstein2023_b} except for those that adopt a top-heavy IMF \citep{Cowley2018,Lu2024} or those with enhanced efficiency of star formation \citep[e.g.][]{McCaffrey2023,spdrv1}. Since dust lowers the observed UV luminosity, if obscuration was the explanation for this galaxy, the intrinsic M$_{\rm UV}$ would be much greater, exacerbating the tension between theoretical models and observations. In contrast, appealing to hotter stars with $T\gtrsim80,000$~K and high gas densities of $\sim10^4\ {\rm cm^{-3}}$ (or even cooler stars at  higher gas density) would also naturally explain this object. Moreover, the inclusion of hotter stars would also help reconcile why the object is so bright. 

However, as discussed previously, the hot/massive star solution implies other behavior. First, we might expect a small UV downturn. At the stellar temperatures considered here, the inferred DLA column density would be $\gtrsim10^{21.8}\ {\rm cm^{-2}}$. Second, depending on the exact shape of the ionizing spectrum, one might expect high EW UV lines. The metal lines can easily be weakened by appealing to low metallicities; however, the key uncertainty is the strength of He~{\small II}~$\lambda$1640. However, the regions of the spectrum at wavelengths below the He$^+$ ionizing threshold is arguably highly uncertain \citep[e.g.][]{Kewley2019}. For this reason, it is unclear how much flexibility there is in the He~{\small II}~$\lambda$1640 predictions. Nevertheless, CEERS2\_588 remains a very viable candidate for hosting hot, metal-poor stars at a high gas density. This same explanation can also be applied to the recently-discovered $z>14$ candidate that has a UV slope of $\sim-2.2\pm0.7$ and very low EW UV lines \citep{Carniani2024} and also GS-z12-0 from \cite{CurtisLake2023,DEugenio2024} at $z=12.6$ with a measured $\beta=-1.84\pm0.19$.

In summary, all but two $z>10$ galaxies are consistent with having strong nebular continuum emission. Most galaxies do not appear to require hot massive stars in very high density gas, but with limited information, we cannot rule out massive stars with normal ISM densities, a top-heavy IMF, or contributions from an extended star formation history. 

\subsection{UV downturns and $\beta$ slopes at $z>10$?}
A key signature of the presence of a strong nebular continuum is the appearance of downturns in the spectrum in the UV, just redward of Ly$\alpha$, or more generally deviations from a power-law slope produced by the stellar continuum. Recently, \cite{Heintz2024} defined the ``Ly$\alpha$ damping parameter'' that encapsulates UV downturns when the value is strongly positive. While Ly$\alpha$ emission can impact this metric, at $z\gtrsim10$, the IGM is likely to absorb the vast majority of Ly$\alpha$ emission and thus, the difference between the measured damping that is expected from the IGM provides a clear probe of the spectral deviation from a pure power-law. 

\cite{Heintz2024} show that the fraction of galaxies with perceived absorption that would imply a neutral column density in excess of $10^{21}\ {\rm cm^{-2}}$ mildly increases from $\sim60\%$ at $z\sim6$ to $65\%-90\%$ at $z>8$. Moreover, \cite{Umeda2023} show that all of their $z>10$ galaxies have UV downturns corresponding to absorption by column densities $>10^{22}\ {\rm cm^{-2}}$. Hence, there is tentative evidence that UV downturns, in particular strong ones, are more common at high-redshift. 

\begin{figure*}
    \centering
    \includegraphics[width=0.45\textwidth]{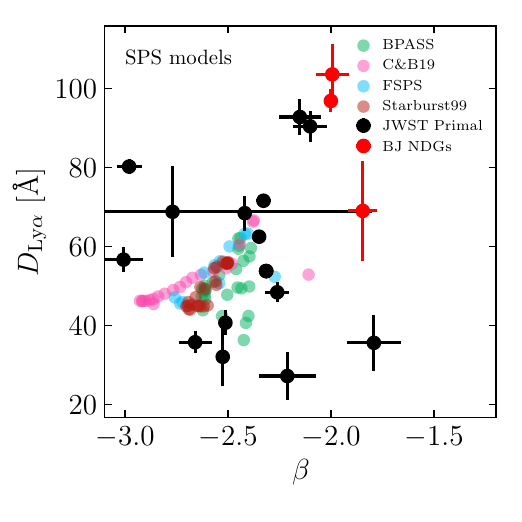}
    \includegraphics[width=0.45\textwidth]{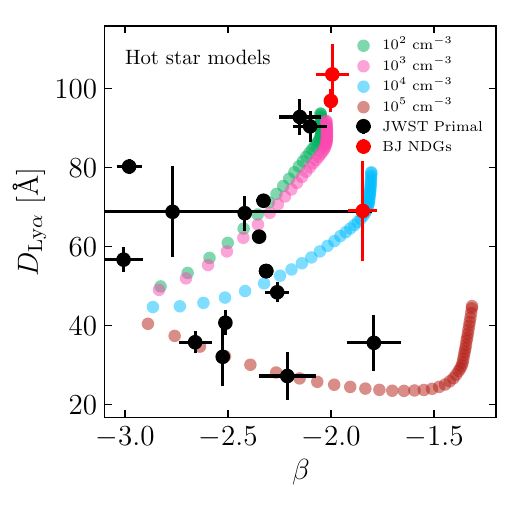}
    \caption{Ly$\alpha$ damping parameter, $D_{\rm Ly\alpha}$, as a function of UV slope compared to various SPS models (left) and for hot metal-poor star models at various gas densities (right). Black data points with error bars represent individual $z>10$ from the JWST Primal database \protect\citep{Heintz2024}. Red points represent the three galaxies in our Balmer jump sample that appear to have a very strong nebular contribution at 1500~\AA. }
    \label{fig:ddla}
\end{figure*}

In Figure~\ref{fig:ddla}, we plot the Ly$\alpha$ damping parameter ($D_{\rm Ly\alpha}$) as a function of UV slope for the various SPS models (left) and the individual metal-poor star models at various ISM densities in the range $10^2-10^5\ {\rm cm^{-3}}$ (right). Note that for this calculation, we have recalculated $\beta$ to match the definition used in \cite{Heintz2024}, although this results in only a marginal change. We also apply IGM absorption assuming the highest redshift curve from \cite{Garel2021}. 

While many of the objects are consistent with the SPS models, two $z>10$ galaxies seem to be outliers for having high $D_{\rm Ly\alpha}\sim90$~\AA\ and a relatively flat UV slope, slightly bluer than $-2.0$, consistent with the galaxies in our Balmer jump sample that seem to exhibit large nebular contribution at 1500~\AA\footnote{Note that to calculate $D_{\rm Ly\alpha}$, we have removed the Ly$\alpha$ line.}. The commonly used SPS models, even with IGM absorption, cannot reproduce these galaxies and hence very dense DLAs would be required as was the interpretation in \cite{Heintz2024,Umeda2023}. However, the right panel of Figure~\ref{fig:ddla} shows that these galaxies are very consistent with the hot star models and a low-density ISM. Since our {\small CLOUDY} calculations were stopped when the electron fraction reached 1\%, no additional absorption is required as the perceived damping is entirely due to H~{\small I} two-photon emission. These galaxies are a $z=12.5$ galaxy from JADES \citep{CurtisLake2023} and a $z=11.4$ galaxy from CEERS \citep{Haro2023}.

Another cluster of high-redshift galaxies appears to have much lower $D_{\rm Ly\alpha}$ and redder $\beta$ than predicted by the SPS models. As we show in the right panel of Figure~\ref{fig:ddla}, these can be reconciled with increasing gas density up to $10^5\ {\rm cm^{-3}}$ (although not necessarily hot massive stars). In this case, two-photon emission is suppressed which reduces the amount of perceived absorption. This can also be reconciled with Ly$\alpha$ emission, though this is rare at these redshifts (although c.f. \citealt{Bunker2023}).

Finally, there is one object (1210\_14220) that has $D_{\rm Ly\alpha}\sim80$~\AA\ and $\beta\sim-3$ which cannot be explained by any of our models. This would imply a dust-free DLA as to not redden the $\beta$, but the DLA must be far enough from the continuum source to keep the nebular emission low enough as to not redden the UV slope. The physical nature of this particular galaxy thus remains uncertain.

\subsection{Considerations for Finding Population III Stars at High-Redshift}
The stellar spectra of hot massive stars are extremely blue \citep[e.g.][]{Schaerer2003} and without a nebular contribution, we expect Pop.~III galaxies to have $\beta<-3.0$. However, the nebular emission is clearly non-negligible and the search for Pop.~III galaxies has primarily consisted of looking for blue ($\beta\lesssim-2$) galaxies that show strong He~{\small II}~$\lambda$1640 emission \citep[e.g.][]{Trussler2023}. However, as discussed in \cite{Cameron2023_BJ} and shown again here (see also \citealt{Trussler2023,Zackrisson2013,Inoue2011}), the two-photon emission is necessarily dominant if the stellar population consists of stars with $T\gtrsim100,000$~K. Hence if strong He~{\small II} emission is found but no UV downturn is present \citep[e.g.][]{Saxena2020}, either the ISM of the galaxy is at very high density (which suppresses the two-photon emission), or the galaxy is not hosting hot massive stars. The latter does not rule out the galaxy from being Pop.~III; however, this would imply that not all Pop.~III stars are extremely hot and might populate a lower range of masses, in agreement with some numerical simulations \citep[e.g.][]{Stacy2016}. In the case of high-densities, the strength of the UV downturn is correlated with the UV slope such that weaker downturns imply redder spectra. We re-emphasize here that the ISM density surrounding Pop.~III stars is unknown. ISM densities surrounding massive stars are typically regulated by a combination of photoheating, radiation pressure, and stellar winds. However in the Pop.~III scenario, winds are much weaker due to the lower opacities of the stellar atmosphere \citep[e.g.][]{Vink2001}, radiation pressure is caused by UV absorption and Ly$\alpha$ scattering rather than IR photons scattering on dust \citep[e.g.][]{Kimm2018}, and finally, the ISM is at much higher pressure due to inefficient cooling \citep[e.g.][]{Omukai2005}, which lowers the pressure gradient between the H~{\small II} region and the PDR. For these reasons, one may reasonably expect that the gas densities around Pop.~III stars may be considerably higher than in the local ISM. We therefore stress that selection criteria for high-redshift Pop.~III galaxies should not rule out systems with red $\beta$ slopes approaching values of $-1$.

\section{Conclusions}
\label{sec:conclusion}
Understanding the physical nature of galaxies at cosmic dawn remains a primary goal of modern cosmology and galaxy formation. Arguably one of the biggest surprises of early high-redshift JWST observations is the number density of both photometrically selected and spectroscopically confirmed objects at $z>10$ as their number counts are significantly higher that expected from many pre-JWST models \citep[e.g.][]{Finkelstein2023_b,Harikane2024,Leung2023,Chemerynska2023}. One of the leading explanations for this anomaly is UV variability driven by extremely bursty star formation histories \citep[e.g.][]{Ren2019,Mason2023,Shen2023,Sun2023,Kravtsov2024}. While most numerical simulations with a resolved ISM predict that star formation histories at early epochs are much burstier than in the low-redshift Universe \citep[e.g.][]{Ma2018,spdrv1}, there remains debate on whether this is enough to reconcile theory with observations \citep{Pallottini2023,Gelli2024}. Hence a detailed study of galaxies with extremely bursty star formation histories is warranted.

One method for probing galaxies with extremely bursty star formation histories is strong nebular emission, and specifically, the visibility of the nebular continuum. More generally, the nebular continuum can significantly modify the observed spectrum of a galaxy, by reddening its spectral slope \citep[e.g.][]{Bouwens2010,Dunlop2013,Cullen2024,Topping2024_beta}, contributing to the 1500~\AA\ luminosity of the galaxy, and causing a downturn in the UV spectrum, just redward of Ly$\alpha$ \citep[e.g.][]{Fosbury2003,Raiter2010,Cameron2023_BJ}. For this reason, we began with a theoretical exercise, adopting five different stellar population synthesis models to quantify how changes in the underlying stellar population synthesis model, IMF slope, stellar temperature, and nebular conditions impact the observed spectra of galaxies under the assumption of an extreme burst of star formation. The conclusions from our theoretical exercise can be summarized as follows:
\begin{enumerate}
\item 1500~\AA\ UV luminosities can increase by up to $\sim0.7$ magnitudes in the presence of extreme bursts of star formation for published stellar population synthesis models at young stellar ages ($\lesssim5$~Myr). This result holds even when considering realistic star formation histories from state-of-the-art numerical simulations. If common at high-redshift, the nebular continuum contribution at 1500~\AA\ has the potential to significantly reduce the amount of ``burstiness'' required to reproduce the measured abundance of bright high-redshift galaxies.
\item Assuming a top-heavy IMF only marginally increases the nebular contribution to the 1500~\AA\ UV luminosity and only slightly reddens the SED as long as the upper-mass limit of the IMF remains fixed. In contrast, assuming hotter (more massive) stars can redden UV slopes to values of $\beta>-2.0$. When the gas density is allowed to increase above the critical density of H~{\small I} two-photon emission, $\beta$ slopes around hot, massive stars can approach $\sim-1$. Hence, searches for Population~III galaxies should not rule out such red objects.
\item When the nebular continuum is strong, two-photon emission can masquerade as DLA absorption with apparent column densities reaching up to $\sim10^{23}\ {\rm cm^{-2}}$. When UV downturns are observed at high-redshift, the DLA solution should be compared with predictions from strong nebular continuum emission.
\item Because $\xi_{\rm ion}$ measurements adopt the observed 1500~\AA\ UV luminosity, observationally inferred $\xi_{\rm ion}$ values deviate from intrinsic $\xi_{\rm ion}$ at values $>10^{25.5}\ {\rm erg^{-1}\ Hz}$. Assuming normal nebular conditions, published SPS models predict that measured values of $\xi_{\rm ion}$ should never be $\gtrsim10^{25.81}$~erg$^{-1}$~Hz. When the effects of dust can be rules out, values above this limit would imply either 1) non-standard ISM conditions, 2) an intrinsic $\xi_{\rm ion}\gtrsim10^{26.2}$~erg$^{-1}$ Hz which would point to objects like hot, massive stars or an AGN, or 3) that the spectrum does not fully cover both the continuum source and the nebular emission.
\end{enumerate}

We then proceeded to study bursty star formation observationally by compiling a sample of 58 galaxies from the Dawn JWST Archive \citep[DJA,][]{Heintz2023} in the redshift interval $2.5<z<9$ that have clear Balmer jumps in their $R\sim100$ JWST NIRSpec prism spectra. This sample increases the number of reported high-redshift JWST Balmer jumps by a factor of $\sim30\times$, confirming some expectations from SED fitting \citep[e.g.][]{Endsley2023}. The conclusions from our observational exercise can be summarized as follows:
\begin{enumerate}
\item SED fitting using non-parametric star formation histories indicates that nearly all sources in the sample are well described by an extreme recent burst of star formation, consistent with expectations for when the Balmer jump becomes visible.
\item Five spectra in the sample are consistent with having a nebular dominated spectrum --- significantly more nebular emission is preferred in the continuum fit to the observed spectrum compared to expectations from SPS models. Two of these spectra come from the same strongly gravitationally lensed galaxy at $z\sim4$ and have red UV slopes, but have Balmer decrements consistent with Case~B recombination expectations. We argue that in these spectra, the MSA slit preferentially covers the nebular gas and not the underlying ionizing source due to the high effective spatial resolution enabled by lensing. In contrast, three other spectra have $\beta\sim-2$, two of which have Ly$\alpha$ emission in addition to a Balmer jump, and the third has an H$\alpha$ EW of $\sim3,300$~\AA. All three have a remarkably similar shape for their UV downturn that is consistent with being two-photon emission from H~{\small I}. The leading explanation is that these galaxies host hot, massive stars as the similarities across multiple galaxies seem to disfavor fine-tuned geometric scenarios that require a DLA.
\end{enumerate}

Finally, we discussed some of the recently detected $z>10$ galaxies in the context of our nebular continuum models. While many galaxies are consistent with expectations for ordinary bursty star formation combined with nebular emission, some seem to deviate by either having excessive downturns in the UV and redder UV slopes. Some of these galaxies are consistent with hot massive star models, which may help explain some of the high-redshift bright galaxy problem.

Because the nebular continuum is sensitive to both ISM density and temperature, a sample such as this provides a unique opportunity to constrain the properties of the high-redshift ISM using emission that has different sensitivities to temperature and density compared to existing strong-line methods. Ideally one could constrain the abundance of Balmer jump galaxies at high-redshift; however, the visual selection and ill-constrained selection functions of the various observational programs used in this work currently inhibits accurate estimates for this quantity. Topics such as this will be explored in future work. 

\section*{Acknowledgments}
HK thanks Andrey Kravtsov for insightful comments and thoughtful discussions. We sincerely thank the PIs and Co-Is of the JWST programs where spectral data was made publicly available on the DJA. We refer interested readers to the following papers for survey descriptions regarding the spectral data: \cite{Bunker2023_spec,DEugenio2024,Bezanson2022,Barrufet2024,Graaff2024,Finkelstein2024,Glazebrook2024,Pierel2024,Siebert2024,Maseda2024}. This work is based in part on observations made with the NASA/ESA/CSA James Webb Space Telescope. The data were obtained from the Mikulski Archive for Space Telescopes at the Space Telescope Science Institute, which is operated by the Association of Universities for Research in Astronomy, Inc., under NASA contract NAS 5-03127 for JWST. These observations are associated with programs listed in Table~\ref{tab:jwstprob}. 
AJC and AS acknowledge funding from the ``FirstGalaxies'' Advanced Grant from the European Research Council (ERC) under the European Union's Horizon 2020 research and innovation programme (Grant agreement No. 789056).

\bibliographystyle{mn2e}
\bibliography{library_oj} 

\appendix
\section{Spectral properties of the nebular continuum}
\label{app:spec_prob}
Our primary analysis has focused on spectral properties of the nebular continuum combined with the transmitted stellar continuum. However, as we have shown in Section~\ref{sec:observations}, a subset of galaxies appear to have spectra with potentially higher nebular fractions than predicted by the various SPS models. Here we discuss the spectral properties of purely nebular spectra. For all calculations we use {\small PyNeb} \citep{Luridiana2015} assuming Case~B recombination, accounting for hydrogen and singly ionized helium with He~{\small I}/H~{\small I}~$=0.08$. A gas density of $10^2\ {\rm cm^{-3}}$ is used for all models.

\subsection{Nebular-only UV slopes}
\begin{figure}
    \centering
    \includegraphics[width=0.33\textwidth]{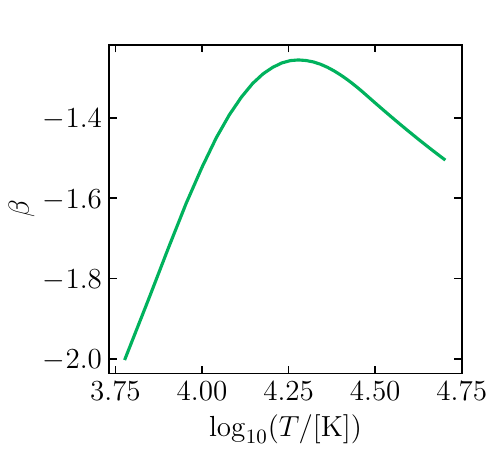}
    \includegraphics[width=0.33\textwidth]{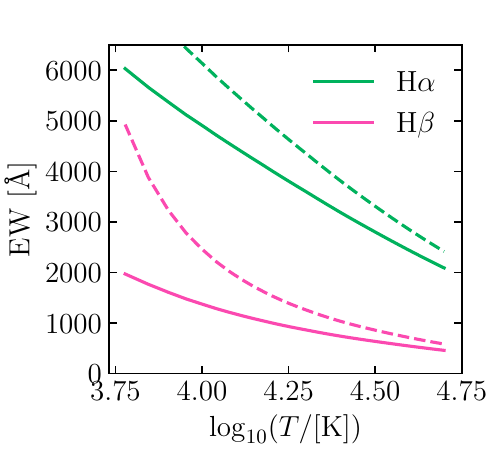}
    \includegraphics[width=0.33\textwidth]{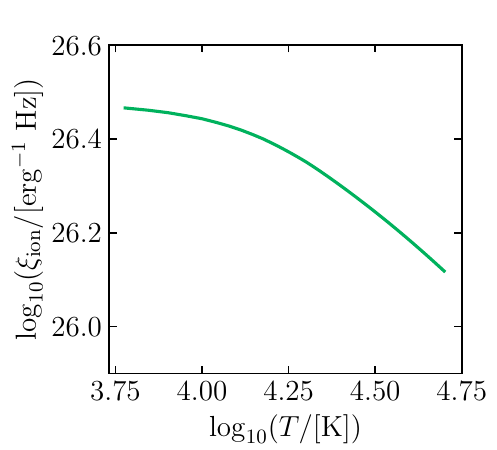}
    \caption{(Left) UV slope, $\beta$ as a function of gas temperature for the nebular continuum assuming Case~B recombination. (Center) Equivalent widths of H$\alpha$ and H$\beta$ as a function of temperature. Solid and dashed lines show the results with and without two-photon emission, respectively. (Right) $\log_{10}(\xi_{\rm ion})$ as a function of gas temperature for the nebular continuum.}
    \label{fig:neb_only_beta}
\end{figure}

In Figure~\ref{fig:neb_only_beta} we show the UV slope, $\beta$, as a function of temperature for the nebular continuum. As above, $\beta$ is measured in the rest-frame wavelength interval $1400\ {\rm \AA}\leq\lambda\leq2600\ {\rm \AA}$. For typical H~{\small II} region temperatures of $T\gtrsim10^4$~K, $\beta$ is redder than $-1.6$, reaching a peak value of $\sim-1.25$. For gas temperatures below $\sim18,000$~K, $\beta$ is a direct measure of gas temperature. However, above this temperature threshold, the temperature corresponding to UV slope becomes double valued. Other indicators would need to be used to elucidate this ambiguity.  

At higher gas densities, where two-photon emission is suppressed, $\beta$ can further redden to strongly positive values but we do not explore this further as this is unlikely for most H~{\small II} regions. 

\subsubsection{Balmer Equivalent Widths}
In the middle panel of Figure~\ref{fig:neb_only_beta} we show the equivalent widths (EWs) of H$\alpha$ and H$\beta$ as a function of temperature. These quantities are strong functions of temperature such that at higher temperatures, the predicted EWs drop significantly. For example, at a temperature of $10,000$~K, the H$\beta$ EW is nearly 1400~\AA, but at $30,000$~K, it drops to 658~\AA. One of the key features of the Balmer line EWs is that they are also sensitive to the presence of two-photon emission. This is perhaps counter-intuitive as the two-photon emission falls off steeply at wavelengths longer than $\sim1450$~\AA\ (in $f_{\lambda}$). The dashed lines in the middle panel of Figure~\ref{fig:neb_only_beta} shows the Balmer line EWs when the two-photon emission has been removed and they are clearly significantly higher in the absence of two-photon emission.

It is important to consider that these calculations all assume Case~B recombination. However, as discussed in \cite{Raiter2010,Ribas2016}, in low-metallicity environments, Case~B departures are expected because the gas is hotter and collisional excitation can preferentially enhance both the Ly$\alpha$ emission and the two-photon emission. The Balmer lines are not enhanced nearly as much due to their higher energies and thus lower collisional excitation coefficients. The effect of these Case~B departures is that the Balmer EWs would then further decrease in low-metallicity environments.

\subsection{Ionizing photon production efficiencies}
In the right panel of Figure~\ref{fig:neb_only_beta} we show the observed $\xi_{\rm ion}$ that would be inferred from H$\alpha$ and $L_{\rm UV,1500\AA}$ as a function of gas temperature for the nebular continuum. This quantity again varies as a function of temperature such that the measured $\xi_{\rm ion}$ decreases with increasing gas temperature. This reflects the decrease in line emissivity with respect to the continuum at high-temperatures. Note that the inferred $\xi_{\rm ion}$ values from a nebular-only spectrum are not too dissimilar from the intrinsic values of the SPS models. We again emphasize that any Case~B departures or cooling radiation would systematically decrease the observed $\xi_{\rm ion}$ from the Case~B values presented here.

\section{Fitting formulae for $\xi_{\rm ion}$ measurements}
\label{app:ff}
Measuring the ionizing photon production efficiency, $\xi_{\rm ion}$ requires converting between an observed emission line (e.g. a Balmer line such as H$\alpha$ or H$\beta$) and the total number of ionizing photons. Typically a single value is used for this conversion that assumes Case~B recombination and a fixed temperature of, for example, $10^4$~K \citep[e.g.][]{Bouwens2016,Matthee2017,Maseda2020}. Under these assumptions the value adopted is $\sim7.3\times10^{11}$~erg$^{-1}$ \citep{Osterbrock2006}. However, as we measure the full posterior distribution on electron temperature from our galaxy fits, we can adopt a temperature-dependent conversion rate. For convenience, we provide fitting formulae to convert H$\alpha$ and H$\beta$ luminosity into ionizing photon luminosity such that:
\begin{equation}
    Q\ [{\rm s^{-1}}] = 10^{-0.007358\log_{10}(T)^2+0.1729\log_{10}(T) + 11.289}\ [{\rm erg^{-1}}] \times L_{\rm H\alpha}\ [{\rm erg\ s^{-1}}]
\end{equation}
and 
\begin{equation}
    Q\ [{\rm s^{-1}}] = 10^{0.0309\log_{10}(T)^2-0.212\log_{10}(T) + 12.679}\ [{\rm erg^{-1}}] \times L_{\rm H\beta}\ [{\rm erg\ s^{-1}}].
\end{equation}
These equations assume that the escape fraction is zero and are based on atomic data from \cite{Pequignot1991}. Note that the ratio of the two conversion factors provides the H$\alpha$/H$\beta$ ratio as a function of temperature. Because H~{\small II} regions in the early Universe tend to be hotter than their low-redshift counterparts, likely due to lower metallicity \citep[e.g.][]{Laseter2024,Morishita2024}, the effect of using a temperature-dependent conversion factor is that the measured $\xi_{\rm ion}$ values can be elevated by $\gtrsim10\%$. 

One final effect to consider is that these conversions assume that all Lyman continuum photons emitted by a source population interact with hydrogen. In practice other species such as helium (and possibly dust) will be present which reduces the ionizing photon budget for hydrogen by $\sim10\%$ \citep[e.g.][]{Tacchella2022_sim}. In this situation, the $Q$ measured from a Balmer line will be an underestimate of the true value.

\end{document}